\documentclass[natbib]{svjour3}  
\smartqed 

\usepackage{natbib}
\usepackage{graphicx}
\usepackage[ruled,linesnumbered,lined,noend]{algorithm2e}
\usepackage{graphicx}
\usepackage{subfigure}
\usepackage{pifont} 
\usepackage{url}
\usepackage{color}
\usepackage{booktabs}
\usepackage{multirow}
\usepackage{latexsym,bm,amsmath,amssymb}
\usepackage{array}
\usepackage{algpseudocode}
\usepackage{lscape}
\usepackage{balance}
\usepackage[colorlinks,linkcolor = blue,citecolor = blue,urlcolor=blue]{hyperref}
\usepackage{tcolorbox}
\usepackage{ragged2e}
\usepackage{mathptmx}      % use Times fonts if available on your TeX system

\definecolor{light-gray}{gray}{0.95}

\newcommand{\tool}{\textsf{SOTitle+}} 
\newcommand{\SOTitle}{\texttt{SOTitle}}

\definecolor{light-gray}{gray}{0.85}
  
\definecolor{mygray}{gray}{.9}

\journalname{Empirical Software Engineering}
\begin{document}

\title{Automatic Bi-modal Question Title Generation for Stack Overflow with Prompt Learning}
% \subtitle{Do you have a subtitle?\\ If so, write it here}

\titlerunning{Stack Overflow Question Title Generation}        % if too long for running head

\author{
	Shaoyu Yang \and Xiang Chen \and Ke Liu \and Guang Yang \and Chi Yu
}

\authorrunning{Yang et al} % if too long for running head

\institute{
Shaoyu Yang \at 
School of Information Science and Technology,\\
Nantong University, Nantong, China\\
\email{shaoyuyoung@gmail.com}
\and
Xiang Chen \at
School of Information Science and Technology\\
Nantong University, Nantong, China\\
\email{xchencs@ntu.edu.cn}
\and
Ke Liu \at
School of Information Science and Technology\\
Nantong University, Nantong, China\\
\email{aurora.ke.liu@outlook.com}
\and
Guang Yang \at 
College of Computer Science and Technology,\\
Nanjing University of Aeronautics and Astronautics, Nanjing, China\\
\email{novelyg@outlook.com}
\and
Chi Yu \at
School of Information Science and Technology\\
Nantong University, Nantong, China\\
\email{yc\_struggle@163.com}
}

\date{Received: date / Accepted: date}
\maketitle
%%%%%%%%%%%%%%%%%%%%%%%%%%%%%%%%%%%%%%%%%%%%%%%%%%%%%%%%%%%%%%%%%%%%%%%%%%%%%%%%%%%%%%%%

\begin{abstract}
When drafting question posts for Stack Overflow, developers may not accurately summarize the core problems in the question titles, which can cause these questions to not get timely help.
Therefore,  improving the quality of question titles has attracted the wide attention of researchers.  
An initial study aimed to automatically generate the titles by only analyzing the code snippets in the question body. 
However, this study ignored the helpful information in their corresponding problem descriptions.
Therefore, we propose an approach {\tool}  by considering bi-modal information (i.e., the code snippets and the problem descriptions)  in the question body. 
Then we formalize the title generation for different programming languages as separate but related tasks and utilize multi-task learning to solve these tasks. Later we fine-tune the pre-trained language model CodeT5 to automatically generate the titles.
Unfortunately, the inconsistent inputs and optimization objectives between the pre-training task and our investigated task may make fine-tuning hard to fully explore the knowledge of the pre-trained model.
To solve this issue, {\tool} further prompt-tunes CodeT5 with hybrid prompts (i.e., mixture of hard and soft prompts).
To verify the effectiveness of {\tool}, we construct a large-scale high-quality corpus from recent data dumps shared by Stack Overflow. Our corpus includes 179,119 high-quality question posts for six popular programming languages.
Experimental results show that {\tool} can significantly outperform four state-of-the-art baselines in both automatic evaluation and human evaluation. In addition, our ablation studies also confirm the effectiveness of component settings  (such as bi-modal information,  prompt learning, hybrid prompts, and multi-task learning) of {\tool}. 
Our work indicates that considering bi-modal information and prompt learning in Stack Overflow title generation is a promising exploration direction.

\keywords{Question Title Generation \and Bi-modal Information \and Code Snippet \and Problem Description \and Prompt Learning \and Multi-task Learning}
\end{abstract}
%%%%%%%%%%%%%%%%%%%%%%%%%%%%%%%%%%%%%%%%%%%%%%%%%%%%%%%%%%%%%%%%%%%%%%%%%%%%%%%%%%%%%%

\section{Introduction}
\label{sec:intro}

Recently, developers widely relied on Stack Overflow (SO) as a primary resource for coding and related information online. With millions utilizing the platform, it serves as a go-to source for finding well-crafted solutions to programming problems. Moreover, SO has evolved into a comprehensive knowledge repository, enabling developers to enhance their programming skills by accessing a wealth of high-quality questions and corresponding answers.
The success of SO largely depends on the user's willingness to answer others' questions. Generally speaking, high-quality questions can increase opportunities for getting assistance. It is also beneficial for developers to find solutions to similar problems, and ultimately for the entire community as it promotes knowledge sharing behavior~\citep{anderson2012discovering,jin2019edits,cao2021automated,gao2023know}.

In support of developers crafting high-quality questions, Stack Overflow has developed comprehensive quality assurance guidelines\footnote{\url{https://stackoverflow.com/help/how-to-ask}} for its community members. However, a significant number of questions posted on SO still fall short of meeting the platform's quality standards.
These low-quality questions lack clarity, precision, or completion, deterring potential experts from providing immediate answers. Consequently, this hampers the advancement and sharing of knowledge within the SO community.
One reason for these low-quality questions is that users cannot draft informative question titles~\citep{correa2013fit,trienes2019identifying,toth2019towards,gao2020generating} as users may not be familiar with the knowledge and terminology related to the problem or have weak writing skills.
Therefore, to improve title quality and reduce the manual efforts for quality maintenance of the SO community, automatic question title generation for Stack Overflow is urgently needed. 
For the convenience of subsequent descriptions, we define \textbf{S}tack \textbf{O}verflow \textbf{Q}uestion \textbf{T}itle \textbf{G}eneration as \textbf{SOQTG}.
For the SOQTG task, we aim to automatically generate titles, which can serve as concise and informative summaries for capturing the essence of a question posted on the Stack Overflow platform.
SOQTG is more challenging as compared to general text summarization~\citep{el2021automatic} or source code summarization~\citep{ahmad2020transformer,iyer2016summarizing,li2022setransformer}.
First, question titles need to be succinct yet informative, capturing the essence of the problem described in the post body.
Second, question titles often involve specialized technical terms and domain-specific language prevalent in software development, making them distinct from general text summarization.
Third,  generating informative titles requires understanding the context provided in the post body and associated code snippets, highlighting the core issue to attract relevant answers.
Finally, question titles should align with community norms and conventions prevalent on the Stack Overflow platform to ensure increased visibility and relevance.

\begin{figure}[htbp]
	\centering
	\includegraphics[width=0.8\textwidth]{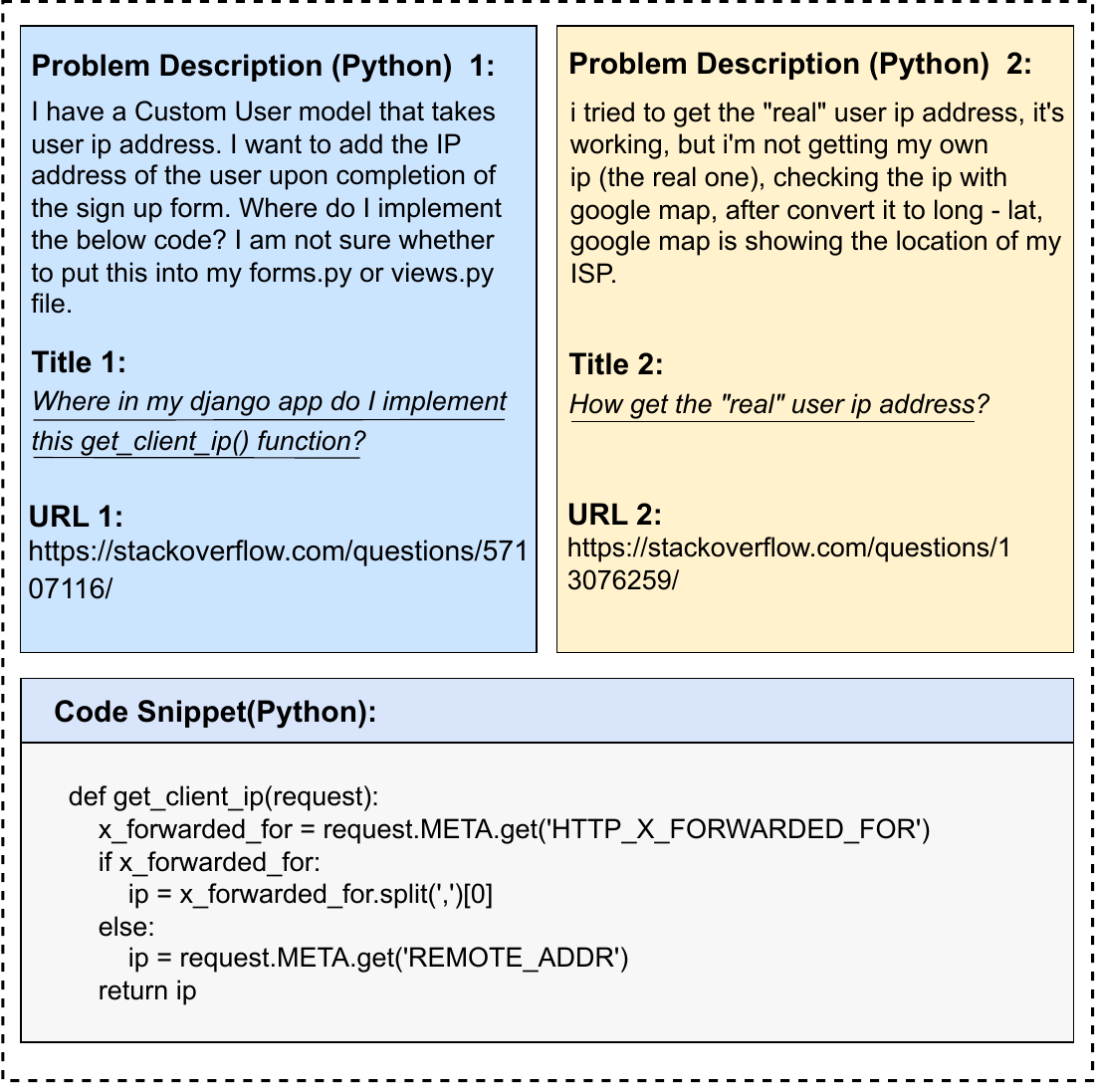}
	\caption{Two posts from Stack Overflow, which have the same code snippet but different problem descriptions}
	\label{fig:motivation1}
\end{figure}

To our best knowledge, \citep{gao2020generating} were the first to automatically generate question titles by analyzing the code snippets in the question body. 
They formalized the SOQTG task as a sequence-to-sequence learning problem and then proposed an approach Code2Que. Specifically,  an attention mechanism is enhanced to perform better content
selection, a copy mechanism is utilized to handle the out-of-vocabulary problem, and a coverage mechanism is used to avoid meaningless repetition.
However, they ignored the valuable information in relevant natural language descriptions, which are called problem descriptions in our study.
As shown in \figurename~\ref{fig:motivation1}, two posts in Stack Overflow have the same code snippet. However, these two posts have different question titles due to different problem descriptions. Based on this motivation example, we find the problem description of the question post can provide valuable information for question title generation, which sometimes cannot be provided by the code snippet. Therefore, considering bi-modal information (i.e., code snippet and problem description) in the question body can help to generate high-quality question titles.

\begin{figure*}[htbp]
	\centering
	\includegraphics[width=1\textwidth]{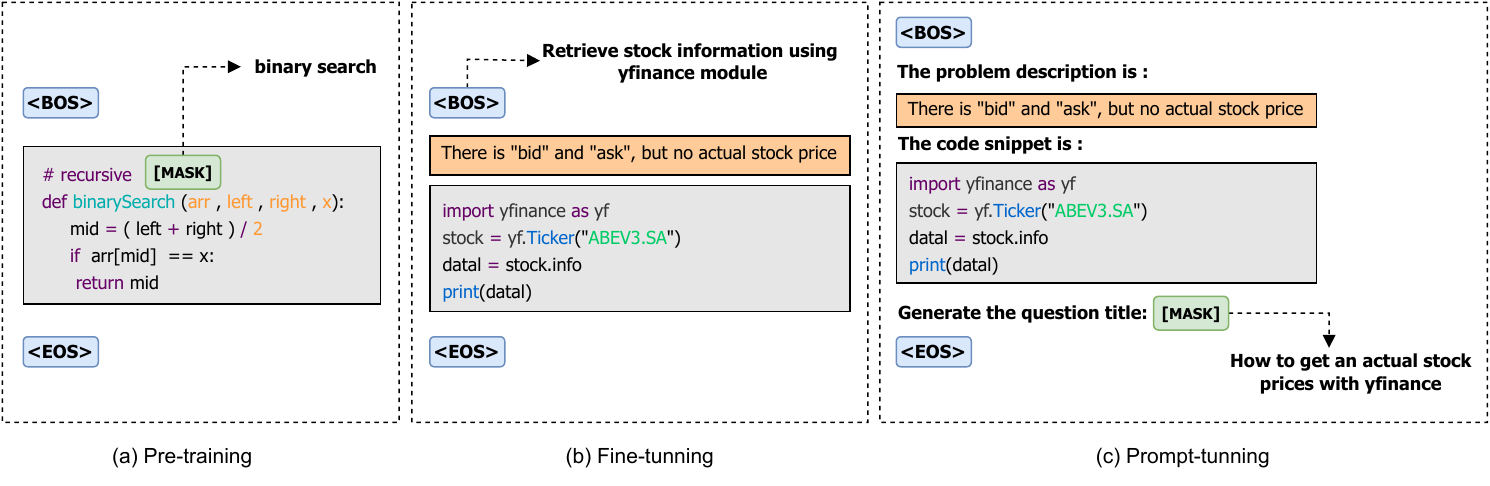}
	\caption{Illustration on the process of pre-training, fine-tuning, and prompt tuning of the SOQTG task. We use [MASK] and [BOS] to denote two special tokens in CodeT5.}
	\label{fig:motivation2}
\end{figure*}

For the SOQTG task, a popular approach is to treat this task as the downstream task and then fine-tune a Pre-trained Language Model (PLM) on the collected corpus of this task.
However, there exist several differences between the pre-training task and the downstream task~\citep{wang2022no}. As shown in \figurename~\ref{fig:motivation2} (a), we use CodeT5~\citep{wang2021codet5}, which is a PLM for code understanding and generation, as an example. 
CodeT5 is pre-trained using the MSP (Masked Span Prediction) optimization objective. 
The input to MSP involves bi-modal information, which is a mixture of code snippets and corresponding natural language descriptions. Then the model is trained to predict randomly masked input tokens. 
However, when CodeT5 is fine-tuned for the downstream task SOQTG,
it loses the special token [\textit{MASK}] and fails to identify the bi-modal information as CodeT5 is trained to a sequence-to-sequence model, which is shown in \figurename~\ref{fig:motivation2} (b). 
Moreover, the optimization objective of the SOQTG task shown in Section~\ref{sec:modelconstruction} is also task-specific.
The inconsistent inputs and optimization objectives between the pre-training task and the downstream task may make fine-tuning hard to elicit the knowledge of the pre-trained model.

Recently, prompt tuning~\citep{liu2023pre} has been introduced to bridge the gap between pre-training and fine-tuning~\citep{schick2020exploiting,lester2021power,wang2022no,huang2022prompt,zhu2023automating}. 
Specifically, prompt tuning aims to explicitly guide the PLM on how to adapt its learned representations for the specific downstream task.
Taking the SOQTG task as an example,
instead of directly mixing bi-modal information for fine-tuning~\citep{liu2022sotitle, zhang2022improving},
% we add some hard prompts before different modal information to help the model distinguish problem description and code snippet. 
% Moreover, prompt-tuning involves rewriting the input by appending soft prompts, which can be trained, such as ``\textit{Generate the question title:} [\textit{MASK}]" at the end of the question body. Subsequently, the model is prompted to predict the masked span [\textit{MASK}]. 
% By utilizing these prompts, prompt-tuning transforms the SOQTG task into a Masked Span Prediction (MSP) problem, bridging the gap between the pre-training stage and the model-tuning stage.
we use prompt tuning to make the model identify bi-modal information by adding additional natural language descriptions. \figurename~\ref{fig:motivation2}(c) demonstrates the concept of prompt-tuning on our investigated SOQTG task in a visual way. In this example, we use two natural language sentences (i.e., ``\textit{The problem description is:}" and ``\textit{The code snippet is:}") to distinguish the problem description and the code snippet. Then, a prompt (such as ``\textit{Generate the question title: [MASK]}") is added after the input sentence, and [\textit{MASK}] is the ground-truth title to be predicted. The PLM predicts the title at the masked token position, as it does in the pre-training stage. Therefore, by adding the prompt, we can reformulate the SOQTG task into an MSP task, aligning the objective with the pre-training stage. This task of reformulation can help fully utilize the task-specific knowledge hidden in the pre-trained model.

We propose the approach {\tool} based on the above research motivation analysis.
Specifically, we first construct the corpus from Stack Overflow. To ensure the quality of the corpus, we use three heuristic selection rules to filter low-quality question posts. 
Since we formalize question title generation for different programming languages as separate but related tasks, we utilize multi-task learning to simultaneously solve these tasks, which can alleviate the insufficient training data issue for low-resource programming languages.
When constructing the model, we use bi-modal information as the input. Moreover, we use hybrid prompts, which is a trade-off between keeping task-related tokens and introducing additional training parameters, to perform prompt-tuning on CodeT5. In the model prediction phase, based on the trained pre-trained model based on CodeT5 and hybrid prompts,  the pre-trained model is prompted to generate the question title with the input of the problem description and code snippet.

To verify the effectiveness of {\tool}, we construct a large-scale high-quality corpus from recent shared data dumps provided by Stack Overflow. 
After using the selection rules,
our corpus contains 179,119 high-quality question posts for six popular programming languages (i.e., Python, Java, C\#, JavaScript, HTML, and PHP). 
In our empirical study,  we compare {\tool} with state-of-the-art baselines via automatic evaluation (i.e., Rouge~\citep{lin2004rouge}, BLEU~\citep{papineni2002bleu}, METEOR~\citep{banerjee2005meteor}, CIDEr~\citep{vedantam2015cider}) and human study. 
Specifically, we first select Code2Que proposed by ~\citep{gao2020generating} as the first state-of-the-art question title generation baseline. 
Then we select two recent question title generation approaches proposed by Zhang et al. (i.e., CCBERT~\citep{zhang2022improving} and M3NSCT5~\citep{zhang2023diverse}) as the other two baselines. Finally, we consider our preliminary work SOTitle~\citep{liu2022sotitle} as the final baseline. 
Final comparison results of both experimental study and human study show that {\tool} can generate higher-quality titles than these four baselines.
Moreover, a set of ablation studies also confirms the effectiveness of using bi-modal information, multi-task learning, prompt tuning, and hybrid prompts in {\tool}.

We extend our preliminary study, which appears as a research paper in SANER 2022~\citep{liu2022sotitle}, as follows:
(1) \textbf{Approach Extension.} Compared with {\SOTitle}~\citep{liu2022sotitle}, We use prompt tuning to replace fine tuning. Specifically, we separate the bi-modal information in the question body with prompt templates, which can help to effectively identify bi-modal information. Then we use the pre-trained model CodeT5~\citep{wang2021codet5} for prompt tuning. Finally, we conduct the ablation study to confirm the effectiveness of using the hybrid prompts in {\tool}.
(2) \textbf{Updated Corpus.} We evaluate the effectiveness of {\tool} by considering recent shared data dumps from Stack Overflow. Moreover, we also consider two more popular programming languages (i.e., HTML and PHP). Finally, We evaluate the performance of the trained model by considering the temporal relationships of posts when we split the constructed corpus. This can guarantee that the test set contains the newest posts and relieve the data leakage problem.
(3) \textbf{More Baselines.} We further consider three baselines, including our preliminary study~\citep{liu2022sotitle} and two other state-of-the-art question title generation approaches CCBERT~\citep{zhang2022improving} and M3NSCT5~\citep{zhang2023diverse}.
(4) \textbf{More Performance Measures.} We further consider three text overlap-based measures to evaluate the model performance: BLEU~\citep{papineni2002bleu}, METEOR~\citep{banerjee2005meteor}, and CIDEr~\citep{vedantam2015cider}.
(5) \textbf{More Discussions.} We first evaluate the effectiveness of {\tool} for two low-resource programming languages (i.e., Ruby and Go). We second compare our approach {\tool} with the Large Language Model (i.e., ChatGPT) on the SOQTG task in the zero-shot scenario. Finally, we analyze the limitations of {\tool} and discuss potential improvement directions for the SOQTG task.

To our best knowledge, the main contributions of our preliminary study and the extensions in this study are summarized as follows:

\begin{itemize}

  \item \textbf{Direction.} We provide two feasible research directions for improving Stack Overflow question title generation. The first direction is to consider multi-modal information (such as problem description, and code snippet) in the question posts. The second direction is to consider prompt learning. Based on the promising performance of our study, we call for more follow-up studies on these two directions for further exploration and advancements in this task.

  \item \textbf{Approach.} We propose a novel approach {\tool} based on the bi-modal information . We apply prompt learning to {\tool} and design the novel hybrid prompts. {\tool} adopts multi-task learning to this task for different programming languages and a pre-trained model CodeT5 is utilized.

  \item \textbf{Corpus.} We construct a high-quality corpus, which contains 179,119 high-quality question posts for six popular programming languages.

  \item \textbf{Evaluation.} We conduct both automatic evaluation and human evaluation to show the competitiveness of {\tool}. Moreover, we conduct a set of ablation studies to show the component setting rationality of {\tool}. 

  \item \textbf{Tool.} Based on {\tool}, we developed a browser tool, which can assist developers in generating question titles for Stack Overflow. We share the video demonstration at YouTube\footnote{\url{https://www.youtube.com/watch?v=_KgUISAT74M}}.
\end{itemize}

\textbf{Open Science.} To support the open science community, we share our scripts, detailed experimental results, and the developed plugin tool in the GitHub repository\footnote{\url{https://github.com/shaoyuyoung/SOTitlePlus}}. Moreover, we also shared our dataset and trained model on Zenodo\footnote{\url{https://zenodo.org/records/10656359}}.

The rest of this paper is organized as follows.
Section~\ref{sec: approach} shows the framework of {\tool} and the details of each phase. 
Section~\ref{sec: setup} shows the experimental setup, including research questions and their design motivation, experimental subject, performance measures, baselines, implementation details, and running platform.
Section~\ref{sec: results} performs a comparison with baselines in the automatic evaluation and human evaluation and conducts a set of ablation studies.
Section~\ref{sec: discuss} discusses the effectiveness of using multi-task learning, question title generation in the low-resource scenario, the comparison with ChatGPT, the limitations of {\tool}, and the main threats to the validity of our empirical study. 
Section~\ref{sec: related} analyzes related work for the SOQTG task, and prompt learning and its applications to software engineering tasks. 
Section~\ref{sec: conclusion} concludes our work and shows potential future directions for SOQTG.

\section{Our Proposed Approach {\tool}}
\label{sec: approach}

\begin{figure*}[htbp]
	\centering
	\includegraphics[width=1\textwidth]{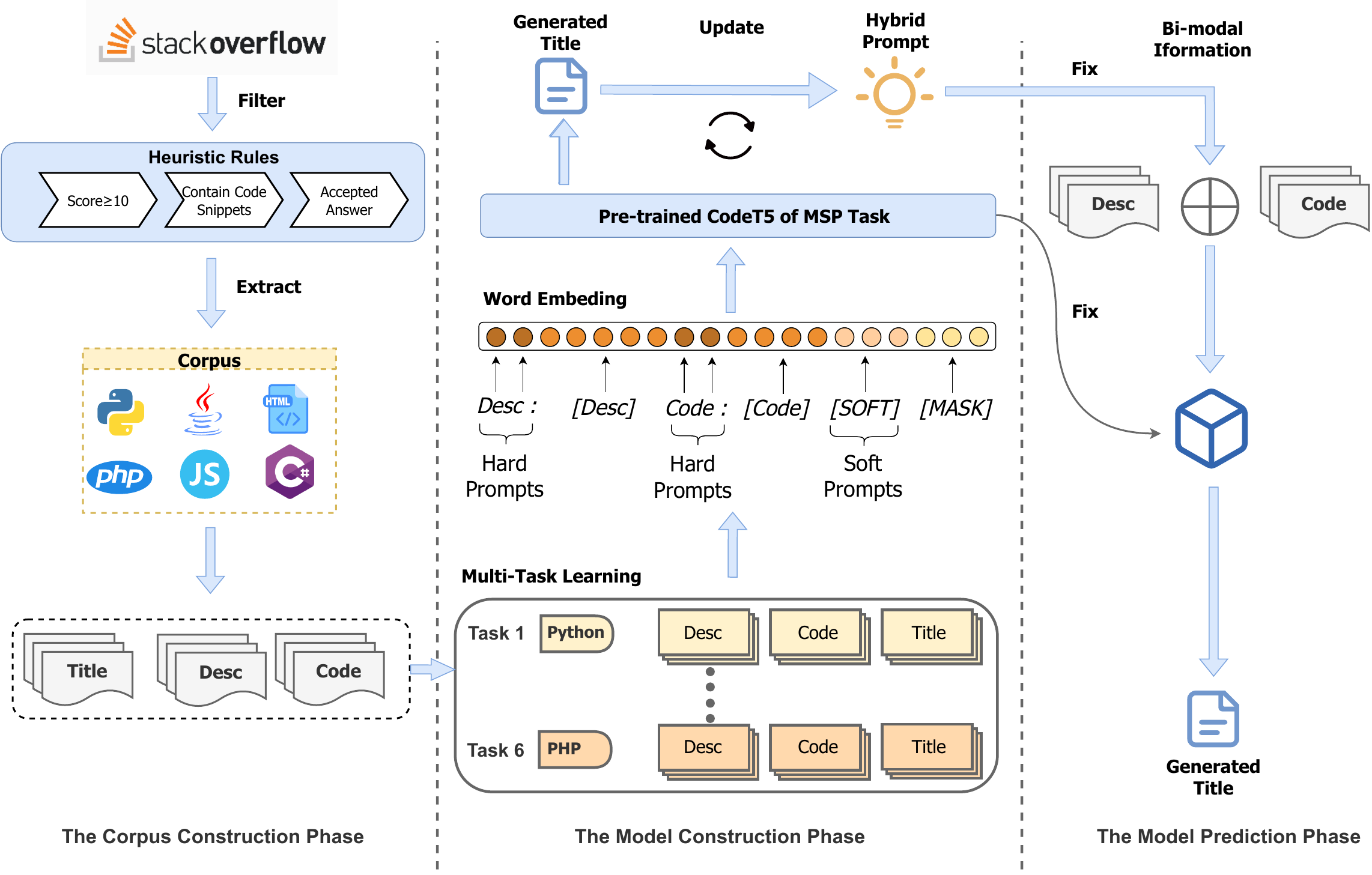}
	\caption{Framework of our proposed approach {\tool}}
	\label{fig:Framwork}
\end{figure*}

The overall framework of  {\tool} is shown in \figurename~\ref{fig:Framwork}.
Specifically, {\tool} consists of three phases: corpus construction phase, model construction phase, and model prediction phase. In the rest of this section, we show the details for each phase.

\subsection{Corpus Construction Phase}
\label{sec:corpusconstruction}

In this phase, we aim to construct the corpus by mining shared data dumps from Stack
Overflow\footnote{\url{https://archive.org/download/stackexchange}, downloaded in March 2023}.
Due to a large number of posts in Stack Overflow, we refer to popularity statistics of tags\footnote{\url{https://stackoverflow.com/tags?tab=popular}, accessed in March 2023} and then determine the top-six popular programming languages (i.e., Python, Java, JavaScript, C\#, PHP, and HTML). 
Moreover, we select two other programming
languages (i.e., Go and Ruby), which can be used to evaluate the effectiveness of {\tool} in the low-resource scenario (discussed in Section~\ref{sec:lowresource}).

\begin{figure}[htb]
	\centering
	\includegraphics[width=0.8\textwidth]{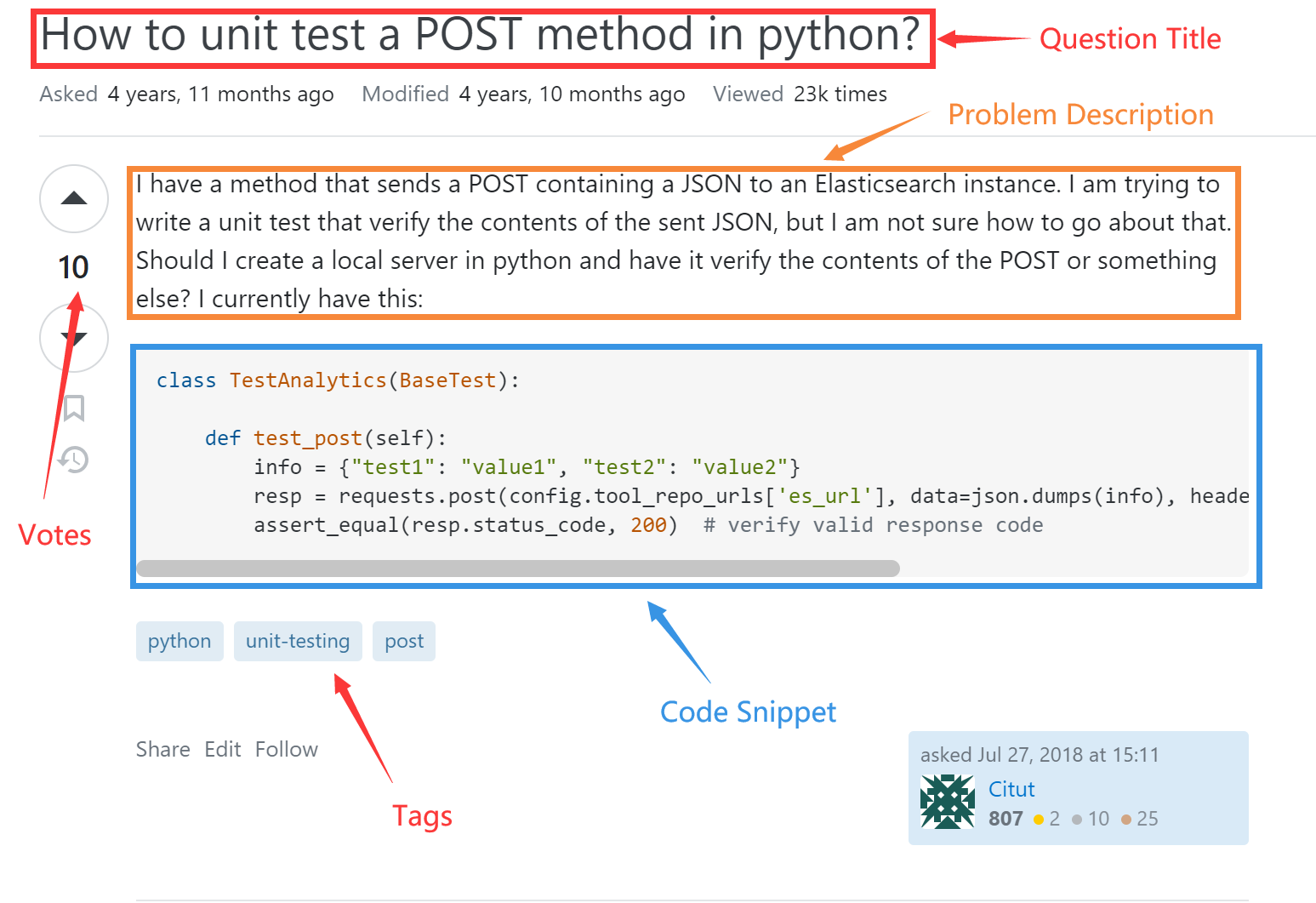}
	\caption{A question post related to Python programming language}
	\label{fig:question_post}
\end{figure}

Since there exist many low-quality posts in our initial gathered corpus, we consider three heuristic selection rules by following previous studies~\citep{islam2019comprehensive,yazdaninia2021characterization,yin2018learning}. 
For the convenience of introducing these selection rules, we use an example post\footnote{\url{https://stackoverflow.com/questions/51560850}} shown in \figurename~\ref{fig:question_post} to illustrate the main components of a Stack Overflow post.
Specifically, 
this question post is related to the Python programming language, which contains a short question title, the problem description with the corresponding code snippet, the number of votes, and multiple tags. After introducing these components, we show the details of these three selection rules and their design motivations as follows.

\begin{itemize}
    \item \textbf{Rule 1:} The votes of the question post should not be smaller than 10. This rule is motivated by the previous study on Stack Overflow post analysis on deep learning bug characteristic analysis~\citep{islam2019comprehensive}. They reduced the number of posts by considering the number of votes to focus on the high-quality posts. Through our preliminary empirical analysis, we found setting this value to 10 can help to ensure a certain level of post quality for the top six different popular programming languages. 
    \item \textbf{Rule 2:} The question post should contain code snippets. The reason for setting this rule is to verify the effectiveness of our approach by considering bi-modal information.
    \item \textbf{Rule 3:} The question post should have an accepted answer. This rule is motivated by previous studies on mining Stack Overflow~\citep{yazdaninia2021characterization,yin2018learning}. We used the existence of an accepted answer as a criterion to potentially indicate question post quality, assuming that such questions might have clearer question titles. 
\end{itemize}

After applying these three selection rules, we select 1.62\% of posts (i.e.,  179,119 posts from 11,023,365 posts).
Then we extract the related programming language, the question title, the problem description, and the code snippet as the quadruplet $\langle$Lang, Title, Description, Code$\rangle$ for each question post. Notice as question posts were stored as a unified HTML format in the data dump, we use the Python Library lxml\footnote{\url{https://lxml.de/}} to extract the text wrapped by the ``$<$code$>$" tag as the code snippet and the remaining text as the problem description.

\subsection{Model Construction Phase}

In this phase, we aim to train the model for question title generation.
Specifically, we first formalize question title generation for different programming languages as separate but related tasks and utilize multi-task learning~\citep{liu2019multi,yang2023syntax} to solve these tasks.
Then we consider the bi-modal information (i.e., problem description and code snippet), which is distinguished by hard prompts  (e.g., ``The problem description" and ``The code snippet"). Meanwhile, we create soft prompts by replacing unimportant tokens with [\textit{SOFT}] tokens.  
Based on the hybrid prompts (i.e., a mixture of the hard prompts and the soft prompts), our model is prompt-tuned based on a pre-trained Transformer model CodeT5~\citep{wang2021codet5}.

\subsubsection{Hybrid Prompt Template Construction}
\label{model-prompt-tuning}

Prompt learning~\citep{liu2023pre} uses the prompt template to modify the original input to generate the new input with a prompt. However, designing a high-quality prompt template for the specific downstream task is still a challenging task.
According to the types of prompt tokens, prompt templates can be classified into three types: hard prompt, soft prompt, and hybrid prompt. 

\textbf{Hard Prompt.} 
A prompt template mainly includes two types of slots: input slot and answer slot.
For the SOQTG task, there are two input slots.
The input slot $[X]$ will be filled with the problem description and the input slot $[Y]$ will be filled with the code snippet.
The answer slot $[Z]$ is eventually filled with the text generated by the PLM.
Except for these input slots and the answer slot, the prompt tokens in the hard prompt~\citep{liu2023pre} are words that humans can understand and are often designed by domain experts. 
Notice that there is no verbalizer for the generation task.
An example of the hard prompt can be found in Fig.~\ref{fig:motivation2} (c).
Therefore, we define the hard prompt for the SOQTG task as follows.

\begin{equation}
\label{equ:hard}
\resizebox{\linewidth}{!}{
    $f_{\text{hard}}= \text{The problem description: }[X] \text{The code snippet: }[Y] \text{ Generate the question title:} [Z]$
}
\end{equation}

\textbf{Soft Prompt.} 
The prompt tokens in the soft prompt~\citep{liu2023pre} are words that humans cannot understand but continuous embeddings that can be tuned during the downstream task training. 
Compared to the hard prompt, the soft prompt can overcome the difficulty of designing appropriate hard prompts for a specific task.
There are two types of soft prompt: vanilla soft prompt and prefix soft prompt~\citep{wang2022no}.
For the SOQTG task, we consider the vanilla soft prompt and define the soft prompt as follows.

\begin{equation}
\label{equ:soft}
f_{\text{soft}} = [\text{SOFT}] \; [X] \; [\text{SOFT}] \;  [Y] \; [\text{SOFT}] \; [Z]
\end{equation}

Notice, these [SOFT] tokens are initialized by using the embeddings of natural words in the hard prompt template in our study.
For example, the third [SOFT] token can be initialized by ``generate the question title:"\footnote{An example used to clarify this process can be found in \url{https://github.com/shaoyuyoung/SOTitlePlus/blob/main/embeddings.md}}.
This setting can ensure the reproduction of our experimental results and help to find the optimal embeddings when compared with the random initialization method.

\textbf{Hybrid Prompts.} 
Hybrid Prompt is a combination of the hard prompt and the soft prompt. Specifically, tokens in the hard prompt are usually important tokens related to the investigated task, which cannot be tuned during the training. On the contrary, tokens in the soft prompt are unimportant and can be tuned during the downstream task training. 
Recently, the effectiveness of using the hybrid prompt has been confirmed in a recent study for vulnerability characteristic prediction~\citep{Li2023Vulnerability}.
For the SOQTG task, we treat the task-related tokens (i.e., ``The problem description:" and ``The code snippet:") as the important tokens, since these tokens can instruct the model to clearly separate the problem description and the code snippet from the bi-modal information. For the remaining tokens (i.e., ``Generate the question title:"), we consider the vanilla soft prompt and replace them with [SOFT] tokens since there are many alternatives to this sentence, such as ``Create a question heading:", ``Draft a post title:" or ``Summarize a question post". Based on the above analysis, we define the hybrid prompt for the SOQTG task as follows.

\begin{equation}
\label{equ:hybrid}
f_{\text{hybrid}} = \text{The problem description: }[X] \;  \text{The code snippet: }[Y] \; [\text{SOFT}] \; [Z]
\end{equation}

\subsubsection{Multi-task Learning}

Multi-task learning~\citep{liu2019multi} is a technique that trains a single model to perform multiple related tasks simultaneously, which can help to improve scalability and efficiency. Multi-task learning has been shown to improve the generalization capabilities of natural language pre-training models by reusing the majority of model weights across multiple tasks~\citep{zhang2017survey}. Since we treat question title generation for different programming languages as separate but related tasks, we prefix the input $X$ of each programming language with a task-specific prefix: [LANG] (e.g., the [LANG] ``JS:" denotes JavaScript) to make the model distinguish different tasks. Then the format of the input can be defined as follows:

\begin{equation}
\mathrm{X}=\text {[LANG]}\oplus \mathrm{X}
\end{equation}

\subsubsection{Prompt Tuning on Pre-trained Model CodeT5}
\label{sec:modelconstruction}

The encoder of CodeT5 is made up of a stack of blocks, each of which mainly consists of two components: a self-attention layer which is followed by a feed-forward network.

Self-attention~\citep{ersulong} is calculated based on a set of queries ($Q$), keys ($K$), and values ($V$). Then dot product is calculated between the queries and keys, Finally, each is divided by $\sqrt{d_{k}}$, and the softmax function is applied to get the weight of the corresponding value.

\begin{equation}
\operatorname{Attention}(Q, K, V)=\operatorname{softmax}\left(\frac{Q K^{T}}{\sqrt{d_{k}}}\right) V
\end{equation}

The encoder includes a fully connected feed-forward network (FFN) at each layer. 
The structure of the decoder is similar to the encoder but with a modification to its self-attention mechanism, which can avoid positions attending to subsequent positions. To facilitate deep architecture, the transformer incorporates residual connections and layer normalization between layers.

To capture subtle semantic differences in the input sequence, multi-head self-attention is employed across different representation subspaces. Multi-head self-attention is defined as follows:

\begin{equation}
\operatorname{MultiHead}(Q, K, V)=\operatorname{Concat}\left(\operatorname{head}_{1},\ldots, \text{head}_{\mathrm{h}}\right) W^{O}
\end{equation}
where $i$-th head $\operatorname{head}_{\mathrm{i}}$ can be calculated as follows:

\begin{equation}
\operatorname{head}_{\mathrm{i}}=\operatorname{Attention}\left(Q W_{i}^{Q}, K W_{i}^{K}, V W_{i}^{V}\right)
\end{equation}
where $W_{i}^{Q}, W_{i}^{K},  W_{i}^{V}$ are linear projection matrices.

Recently, ~\citep{niu2023empirical} showed that using CodeT5~\citep{wang2021codet5} could achieve state-of-the-art performance in many downstream tasks (such as source code summarization and code translation) for software engineering. CodeT5 was proposed by ~\citep{wang2021codet5}, which was trained on a large-scale corpus consisting of bi-modal information (NL-PL). Specifically, Wang et al. randomly masked spans with arbitrary lengths and then predicted these masked spans combined with some sentinel tokens at the decoder, They referred to this task as \textbf{Masked Span Prediction (MSP)}, as illustrated in \figurename~\ref{fig:motivation2}(a). The loss function of MSP is defined as follows:

\begin{equation}
\mathcal{L}_{M S P}(\theta)=\sum_{t=1}^{k}-\mathrm{log}\,{P}_{\theta}(x_{t}^{\mathrm{mask}}|\mathbf{x}^{\mathrm{\backslash mask}},{\bf x}_{<t}^{\mathrm{mask}})
\end{equation}
where $\theta$ denotes the model parameters, $\mathbf{x}^{\mathrm{\backslash mask}}$ denotes the masked input, $\mathbf{x}^{\mathrm{mask}}$ denotes the masked sequence to predict from the decoder with $k$ denoting the number of tokens in $\mathbf{x}^{\mathrm{mask}}$, and ${\bf x}_{<t}^{\mathrm{mask}}$ is the span sequence generated so far.

Since we utilize multi-task learning for processing these six different programming languages, we define the final loss function as follows:

\begin{equation}
\mathcal{L}_{LOSS}(\theta)=(\sum_{i=1}^{6}\mathcal{L}_{M S P}^{Task_i}(\theta))/6
\end{equation}
where each $Task_i$ represents different tasks (i.e., different programming languages).

\subsection{Model Prediction Phase}

In this phase,  
for the problem posts that require title generation, we extract the problem description and the code snippet from the post body. Subsequently, based on the tuned hybrid prompt, we can construct the input and feed it into the PLM CodeT5 to obtain the generated question title.

\section{Experiment Setup}
\label{sec: setup}

\subsection{Research Questions}

In our empirical study, we want to answer the following five research questions (RQs):

\textbf{RQ1:} Can our proposed approach {\tool} generate higher-quality question titles than four state-of-the-art baselines via automatic evaluation?

\textbf{Motivation.} To show the competitiveness of {\tool} when compared to the baselines, we consider four performance measures to compare the similarity between the generated question title and the ground-truth question title in this RQ.

\textbf{RQ2:} Whether using the bi-modal information in the question body can improve the performance of {\tool}?

\textbf{Motivation.} In this RQ, we want to conduct the ablation study to investigate which input modal contributes the most to the performance of {\tool}.

\textbf{RQ3:} Whether using the prompt-tuning paradigm can improve the performance of {\tool} than using the fine-tuning paradigm?

\textbf{Motivation.}  In this RQ, we want to conduct the ablation study to investigate whether using the prompt-tuning paradigm can achieve better performance of {\tool} than using the fine-tuning paradigm.

\textbf{RQ4:} Whether using hybrid prompts can improve the performance of {\tool}?

\textbf{Motivation.} Prompt template design is a challenging and open problem in both natural language processing and software engineering tasks. In this RQ, we want to conduct the ablation study to show the competitiveness of using hybrid prompts in our proposed approach {\tool}.

\textbf{RQ5:} Can our proposed approach {\tool} outperform state-of-the-art baselines via human study?

\textbf{Motivation.} There exist some disadvantages to automatic evaluation measures. For example, the text overlap-based measures used in our study may not effectively reflect the similarity between the generated question title and the ground-truth question title from a semantic perspective~\citep{hu2020deep}. Therefore, we want to conduct a human study to evaluate the competitiveness of {\tool} in this RQ. 

\subsection{Experimental Subject}
\label{sec4:subject}

After the corpus construction phase (introduced in Section~\ref{sec:corpusconstruction}),
we finally select 43,056 posts for Python, 34,297 posts for Java, 37,900 posts for JavaScript, 34,846 posts for C\#, 14,288 posts for HTML, and 14,732 posts for PHP. In the process of our corpus split, we consider the temporal relationship of posts. 
Specifically, splitting posts in chronological order is more applicable to the real-world application scenario and this setting can guarantee that the test set contains the latest posts. Therefore, this corpus split setting can relieve the data leakage problem caused by the homogeneous questions between the training set and the test set. Finally, we split the gathered corpus as the training set, the validation set, and the testing set according to 80\%: 10\%: 10\%  respectively. 
In Table~\ref{tab:CORPORA}, we show the detailed statistical information of our corpus split results. 
Finally, we also show the length statistics of code snippets, problem descriptions, and question titles in  Table~\ref{tab:DatasetStatistics}. 
Notice, some posts do not contain any problem description\footnote{\url{https://stackoverflow.com/questions/1478248}}, so the minimum value of the description length is 0.

\begin{table}[htbp]
  \centering
  \caption{Statistical information of our corpus split results}
  \resizebox{0.6\textwidth}{!}{
    \begin{tabular}{ccccc}
    \toprule
    \textbf{Language} & \textbf{Training Set} & \textbf{Validation Set} & \textbf{Testing Set} & \textbf{Total} \\
    \midrule
    Python  & 34,444 & 4,306 & 4,306 & 43,056\\
    Java\   & 27,437 & 3,430 & 3,430 & 34,297 \\
    JavaScript & 30,320 & 3,790 & 3,790 & 37,900\\
    C\# & 27,876 & 3,485 & 3,485 & 34,846\\
    HTML & 11,430 & 1,429 & 1,429 & 14,288\\
    PHP & 11,785 & 1,473 & 1,474 & 14,732\\
    \midrule
    \textbf{Total} & \textbf{143,292} & \textbf{17,913} & \textbf{17,914} & \textbf{179,119}\\
    \bottomrule
    \end{tabular}}%
  \label{tab:CORPORA}%
\end{table}%

\begin{table*}[htbp]
 \caption{Length statistics of code snippets, problem descriptions, and question titles in our corpus}
 \begin{center}
 \setlength{\tabcolsep}{1mm}{
   % \resizebox{1.0\textwidth}{}{
    \begin{tabular}{c|c|cccccc}
    \toprule
    \textbf{Length Metrics} & \textbf{Values} & \textbf{Python}  & \textbf{Java}  & \textbf{C\#}
    &\textbf{JavaScript}    
    &\textbf{HTML}  &\textbf{PHP} \\   
    \midrule
    \multirow{4}{*}{Code Length}
    & Maximum & 17,629 & 27,348 & 18,575 & 17,568  & 17,568 & 18,192   
    \\
    & Average & 190.82 & 271.49 & 165.07 & 164.95 & 173.23 & 174.89 
    \\
    & Mode & 29 & 17 & 27 & 40 & 20 & 22  
    \\
    & Median & 92 & 96 & 86 & 86 & 86 & 79 
    \\
    & Minimum & 2 & 2 & 2 & 2  & 2 & 2  
    \\
    & \textless{}256 & 81.90\% & 77.13\% & 83.99\% & 84.37\% & 83.97\% & 84.12\% 
    \\
    \midrule
    \multirow{4}{*}{Description Length}
    & Maximum & 1,776 & 2,158 & 1,851 & 1,579  & 1,360 & 1,139  
    \\
    & Average & 87.99 & 94.81 & 101.14 & 86.38 & 83.12 & 86.37 
    \\
    & Mode & 41 & 50 & 49 & 40 & 33 & 46  
    \\
    & Median & 70 & 73 & 77 & 68 & 66 & 66 
    \\
    & Minimum & 0 & 0 & 0 & 0  & 0 & 0  
    \\
    & \textless{}256 & 97.19\% & 95.93\% & 94.85\% & 97.87\% & 97.47\% & 96.57\%
    \\
    \midrule
    \multirow{4}{*}{Title Length}
    & Maximum & 56 & 60 & 50 & 45  & 45 & 49  
    \\
    & Average & 8.56 & 8.48 & 8.61 & 8.42 & 8.55 & 8.29
    \\
    & Mode & 7 & 8 & 8 & 7 & 7 & 7 
    \\
    & Median & 8 & 8 & 8 & 8 & 8 & 8
    \\
    & Minimum & 2 & 2 & 2 & 2  & 2 & 2  
    \\
    & \textless{}16 & 96.63\% & 96.56\% & 95.70\% & 97.06\% & 96.65\% & 97.35\% 
    \\
    \bottomrule
\end{tabular}}
 \end{center}
 \label{tab:DatasetStatistics}
\end{table*}

\subsection{Performance Measures}

To show the competitiveness of {\tool},
we consider four performance measures: Rouge-L~\citep{lin2004rouge}, BLEU~\citep{papineni2002bleu}, METEOR~\citep{banerjee2005meteor}, and CIDEr~\citep{vedantam2015cider}.
Using these measures, we can measure the lexical similarity between the generated question title and the ground-truth question title. Moreover, these measures have been widely used in previous text generation tasks~\citep{lin2023gen,liu2019generating,li2022setransformer,raffel2019exploring,gros2020code,yang2023exploitgen,yang2022ccgir,li2021secnn}.

  \begin{itemize}
      \item \textbf{ROUGE-L.} ROUGE~\citep{lin2004rouge} is a recall-based measure, which is used to calculate the lexical overlap between the generated sentence and the ground-truth sentence. In our study, we use ROUGE-L, which is based on the LCS (Longest Common Subsequence).
      
      \item \textbf{BLEU.} BLEU~\citep{papineni2002bleu} is employed to assess the similarity degree between the generated sentence and the ground-truth sentence. Specifically, BLEU-$n$  calculates the number of \textit{$n$-grams} that appear in the generated sentence.
      
      \item \textbf{METEOR.} METEOR~\citep{banerjee2005meteor} has been widely used to measure the quality of machine translation. It incorporates many aspects (e.g., vocabulary, grammar, and semantics) to evaluate the quality of the generated sentences.
      
      \item \textbf{CIDEr.} CIDEr~\citep{vedantam2015cider} is a measure that evaluates the quality of the generated sentences by analyzing the frequency of $n$-grams in the ground-truth sentences. To accomplish this goal, CIDEr employs TF-IDF weighting for each $n$-gram. Moreover, CIDEr calculates the $CIDEr_n$ score for each $n$-gram by averaging the cosine similarity between the generated sentence and the ground-truth sentence.
  \end{itemize}

Notice the higher the value of these performance measures, the better the performance of the corresponding approach.
In our study, we use the implementations provided by nlgeval\footnote{\url{https://github.com/Maluuba/nlg-eval}} to calculate the value of these four performance measures.

\subsection{Baselines}

For the Stack Overflow question title generation, we select four state-of-the-art approaches as our baselines to show the competitiveness of {\tool}.  
We briefly introduce the characteristics of these baselines as follows.

\begin{itemize}
    \item \textbf{Code2Que.} Code2Que is the first study for Stack Overflow question title generation, which was proposed by ~\citep{gao2020generating}.  Code2Que automatically generates question titles for Stack Overflow posts through an LSTM-based deep learning approach. The model's encoder is a multi-layer, bidirectional LSTM network that processes the input code tokens sequentially. While the decoder is a single-layer LSTM that recursively generates predicted tokens.
    
    \item \textbf{SOTitle.} To show the effectiveness of {\tool}, we also consider our previously proposed approach SOTitle~\citep{liu2022sotitle} as the baseline. SOTitle utilizes the pre-trained model T5~\citep{raffel2019exploring}, which follows a Transformer encoder-decoder architecture and employs multi-task learning to unify text-based language problems into a text-to-text paradigm. 
    
    \item \textbf{CCBERT.} CCBERT~\citep{zhang2022improving} is another approach, which also considers bi-modal information. CCBERT considers the CodeBERT model and introduces customized copy attention layers on the encoder and decoder.

    \item  \textbf{M3NSCT5.} M3NSCT5~\citep{zhang2023diverse} is a recently proposed approach that can improve the quality and diversity of the question titles generated from the code snippets. Zhang et al. fine-tuned the codeT5 and utilized the nucleus sampling and maximal marginal ranking strategy to make the model return a set of high-quality and diverse title candidates.
  
\end{itemize}

To alleviate the internal threats, we utilize the scripts shared by these four baselines and follow the hyperparameter settings used in their original studies.
Since the baselines Code2Que and M3NSCT5 only consider code snippets as their inputs, we still use bi-modal information as their inputs to guarantee a fair comparison. 

\subsection{Implementation Details}

In our empirical study, we use Pytorch library\footnote{\url{https://pytorch.org/}} and transformers library\footnote{\url{https://github.com/huggingface/transformers}}  to construct our model. 
For prompt tuning, we use OpenPrompt library\footnote{\url{https://github.com/thunlp/OpenPrompt}}.
Specifically, by following the tutorial for OpenPrompt~\citep{ding2022openprompt}, we use the \textit{manual\_template}, \textit{SoftTemplate}, \textit{MixedTemplate} APIs to construct the hard prompt, the soft prompt, and the hybrid prompt templates, respectively. Based on the constructed prompt templates, we use \textit{PromptForGeneration} API to perform prompt tuning. More details about our implementation can be found in our shared 
GitHub repository. 

For the model hyperparameter setting, the word embedding dimension and hidden size are set to 768, and the number of attention heads and layers is set to 12.
All parameters were optimized with AdamW~\citep{loshchilov2018fixing}, and the initial learning rate is set to 0.00005.
During training, the batch size is set to 16. The maximum length of the encoder and decoder is set to 512 and 64 respectively.
To implement heuristic search, we adopt beam search and set the beam size to 10. Finally, we also use the early stop method~\citep{prechelt1998early} to further alleviate the overfitting problem, and the weights with the highest performance on the validation set are taken as the final weights of the neural network.

\subsection{Running Platform and Model Training Time}

We run all the experiments on a computer with an Intel(R) Core(TM) i5-13600K and a GeForce RTX4090 GPU with 24 GB memory. 
The running operation system platform is Windows Operation System.
Based on our platform, we need about 24 hours to train the model by using {\tool}.
\section{Result Analysis}
\label{sec: results}

\subsection{RQ1: Comparison with Baselines via Automatic Evaluation}
\label{sec:resultRQ1}

The comparison results between  {\tool} and baselines can be found in Table~\ref{tab:RQ1}. In this table, we emphasize the best value in bold for each column. 
Based on the comparison results, we first find that our proposed approach {\tool} can outperform four baselines in terms of the performance measure Rouge-L for different programming languages.
In particular, for our considered baselines, we find fine-tuning-based baselines (i.e., CCBERT~\citep{zhang2022improving}, SOTitle~\citep{liu2022sotitle} and M3NSCT5~\citep{zhang2023diverse}) can achieve better performance than Code2Que~\citep{gao2020generating}, which trains an LSTM model from scratch. The potential reason is that these three baselines are based on PLMs, which have been trained on the large-scale corpus. Moreover, CCBERT and SOTitle achieve considerable performance, which is better than M3NSCT5. We attribute the good performance of CCBERT and SOTitle to the reason that these two baselines were specifically designed to process bi-modal information.
Finally, our approach can outperform the fine-tuning-based baselines. Specifically,  compared to the baseline CCBERT, {\tool} can improve the performance of ROUGE-L by 18.04\%, 14.65\%, 9.41\%, 13.43\%, 20.19\% and 7.98\% for Python, Java, C\#, JavaScript, PHP, and HTML respectively. 
Moreover, we also perform the Wilcoxon signed-rank test~\citep{wilcoxon1992individual} at the confidence level of 95\% to check whether the performance differences between {\tool} and baselines are significant. All the $p$-values are smaller than 0.05, which means {\tool} can significantly outperform the baselines.

Except for Rouge-L performance measure, we also compare {\tool} with baselines in terms of METEOR~\citep{banerjee2005meteor}, BLEU~\citep{papineni2002bleu} and CIDEr~\citep{vedantam2015cider} performance measures. Final comparison results and the Wilcoxon signed-rank test results also show the competitive performance of {\tool} (i.e., the performance improvement is statistically significant). 

\begin{table*}[htb]
  \centering
  \caption{Comparison results between our proposed approach {\tool}  and state-of-the-art baselines }
  \begin{center}
    \begin{tabular}{c|c|ccccc}
    \toprule
    \textbf{Language} & \textbf{Approach} & \textbf{ROUGE-L (\%)}  & \textbf{METEOR (\%)}  & \textbf{BLEU-1 / 2 / 3 / 4 (\%)}
    &\textbf{CIDEr } \\   
    \midrule
    \multirow{5}{*}{Python}
    & Code2Que & 21.75 & 14.87 & 23.80 /  15.54  /  10.92 / \hspace{0.1em}  8.10 &  0.75  
    \\
    & M3NSCT5 & 25.39 & 17.84 & 27.72  /  19.26  /  14.35  /  11.20   &  0.98   
    \\
    & CCBERT & 26.50 & 15.52 &  28.07  /  19.09  /  13.94  /  10.63   &  1.03  
    \\
    & SOTitle & 26.49 & 17.87  & 26.75 / 19.31 / 14.89 / 11.98 & 1.11  
    \\
    & \textbf{SOTitle+} & \textbf{31.27} & \textbf{22.00} & \textbf{33.28 / 25.03 / 19.98 / 16.59}  & \textbf{1.43} & 
    \\
    \midrule
    \multirow{5}{*}{Java}
    & Code2Que & 19.67 & 13.77 & 23.80 / 15.54 / 10.92 / \hspace{0.1em} 7.14  & 0.66  
    \\
    & M3NSCT5 & 22.83 & 16.46 & 25.43 / 17.37 / 12.70 / \hspace{0.1em} 9.89  & 0.86   
    \\
    & CCBERT & 24.17 & 14.33 & 26.11 / 17.68 / 12.90 / \hspace{0.1em} 9.93  & 0.91     
    \\
    & SOTitle & 23.70  & 16.14  & 24.10 / 17.34 / 13.32 / 10.68 & 0.99  
    \\
    & \textbf{SOTitle+} & \textbf{27.71} & \textbf{19.46} & \textbf{30.04 / 21.99 / 17.18 / 14.01}  & \textbf{1.21} & 
    \\
    \midrule
    \multirow{5}{*}{C\#}
    & Code2Que & 20.00 & 14.29 & 21.19 / 13.71 / \hspace{0.1em} 9.43 / \hspace{0.1em} 6.94  & 0.64  
    \\
    & M3NSCT5 & 23.45 & 16.83 & 24.75 / 17.09  / 12.70 / \hspace{0.1em} 9.98  & 0.85
    \\
    & CCBERT & 25.28 & 15.32 & 27.84 / 19.87 / 14.86 / 11.40  & 0.94 
    \\
    & SOTitle & 24.22  & 16.39  & 25.20 / 18.31 / 14.14 / 11.43 & 0.97  
    \\
    & \textbf{SOTitle+} & \textbf{27.66} & \textbf{19.37} & \textbf{30.69 / 22.94 / 18.27 / 15.24}  & \textbf{1.20} 
    \\
    \midrule
    \multirow{5}{*}{JavaScript}
    & Code2Que & 22.40 & 15.45 & 24.18 / 15.75 / 11.22 / \hspace{0.1em} 8.51  & 0.77 
    \\
    & M3NSCT5 & 25.97 & 18.11 & 27.95 / 19.58 / 14.92  /  11.97   & 1.02
    \\
    & CCBERT & 27.70 & 16.23 & 29.50  / 20.50 / 15.34  /  12.05   & 1.12 
    \\
    & SOTitle & 27.19  & 18.13  & 27.56 / 20.18 / 15.81 / 12.93 & 1.15
    \\
    & \textbf{SOTitle+} & \textbf{31.42} & \textbf{21.90} & \textbf{33.69 / 25.29 / 20.27 / 16.91}  & \textbf{1.41} 
    \\
    \midrule
    \multirow{5}{*}{PHP}
    & Code2Que & 22.50 & 15.63 &  24.17  / 15.60 / 11.22  / \hspace{0.1em}  8.51   &  0.82  
    \\
    & M3NSCT5 & 26.32 & 18.68  & 28.53 / 19.82  / 14.95  / 11.83  & 1.10 
    \\
    & CCBERT & 26.60 & 16.26 & 28.39 / 19.94 /  15.21 / 12.06   &  1.09 
    \\
    & SOTitle & 27.69  & 19.25  & 28.63 / 21.07 / 16.45 / 13.35 & 1.23 
    \\
    & \textbf{SOTitle+} & \textbf{31.97} & \textbf{23.20} & \textbf{34.53 / 25.80 / 20.82 / 17.59}  & \textbf{1.51}
    \\
    \midrule
    \multirow{5}{*}{HTML}
    & Code2Que & 22.80 & 16.67 &  23.39  /  14.87  /  10.11  /  \hspace{0.1em} 7.16   &  0.79 
    \\
    & M3NSCT5 & 25.06 & 18.41 &  25.96  /  16.81  /  11.65  /  \hspace{0.1em} 8.39   &  0.90 
    \\
    & CCBERT & 26.91 & 16.98 &  29.38  /  19.59  /  13.46  / \hspace{0.1em} 9.54   &  0.98 
    \\
    & SOTitle & 25.84  & 18.54  & 28.36 / 18.83 / 13.37 / \hspace{0.1em} 9.76 & 0.98 
    \\
    & \textbf{SOTitle+} & \textbf{29.00} & \textbf{21.03} & \textbf{30.74 / 20.82 / 15.11 / 11.51}  & \textbf{1.14} 
    \\
    \bottomrule
\end{tabular}
 \end{center}
  \label{tab:RQ1}
\end{table*}%

\begin{figure*}[htb]
	\centering
	\includegraphics[width=1\textwidth]{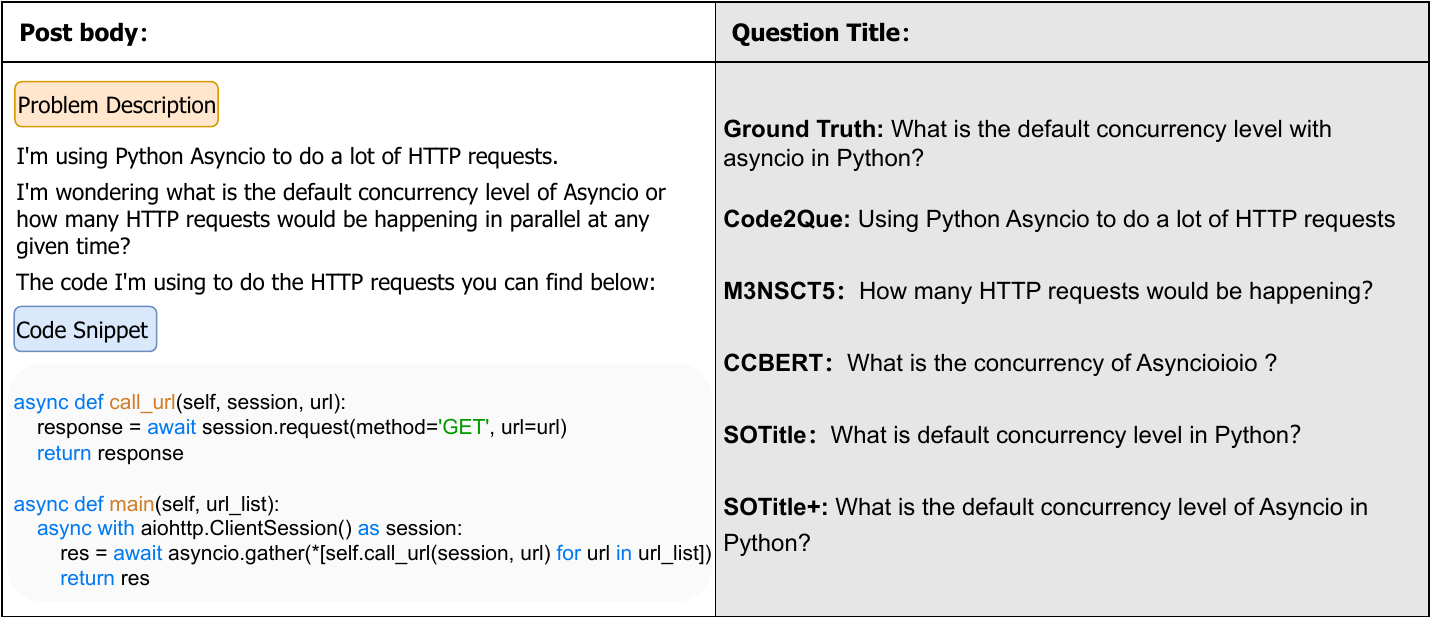}
	\caption{The question titles generated by {\tool} and baselines for a question post related to the Python programming language}
	\label{fig:genCompare}
\end{figure*}

\figurename~\ref{fig:genCompare} shows the question titles generated by {\tool} and baselines for a question post related to Python programming language\footnote{\url{https://stackoverflow.com/questions/66287470}}.
In this example, we can find
 that the question titles generated by Code2Que and M3NSCT5 fail to capture the developer's intent, and they mislocate key information to the ``\textit{HTTP request}".  
 The question titles generated by  CCBERT and SOTitle lose the keywords ``\textit{default}" and ``\textit{Asyncio}", respectively. However, the question title generated by {\tool}  can accurately and fluently express the key information in this post.

\begin{tcolorbox}[width=1.0\linewidth, title={}]
\textbf{Summary for RQ1:} 
 {\tool} can achieve better performance than four state-of-the-art baselines for different programming languages via automatic evaluation.
\end{tcolorbox}

\subsection{RQ2: Ablation Study on Using the Bi-modal Information}

In this RQ, we aim to investigate the contribution of different modal information to the performance of {\tool}. 
Here we use different subscripts to distinguish different control approaches and introduce the meaning of these subscripts as follows.

\begin{itemize}
    \item \textbf{code.} The corresponding control approach only uses the code snippet along with its hard prompt as the input modality. We use {$w/o \ desc$} to denote this control approach.
    \item \textbf{desc.} The corresponding control approach only uses the problem description along with its hard prompt as the input modality. We use {$w/o \ code$} to denote this control approach.
\end{itemize}

We show the performance of {\tool} with different modal information in Table~\ref{tab:RQ2}. When only focusing on code snippets ($w/o \ desc$), the performance of {\tool} causes a significant decrease. In particular, in terms of Rouge-L, the average performance of {\tool} drops by 51.85\%. On the other hand, when only focusing on problem descriptions ($w/o \ code$), there is a slight decrease in the performance of {\tool}. These findings show that problem descriptions can provide more assistance in improving the performance of {\tool} than code snippets. Moreover, using both problem descriptions and code snippets simultaneously can help to achieve the best performance due to the complementarity between these two different modal information.

\begin{tcolorbox}[width=1.0\linewidth, title={}]
\textbf{Summary for RQ2:}
Problem description contributes more to the performance of {\tool} when compared to code snippet. Moreover, Considering both of these two different modal information can achieve the best performance for {\tool}.
\end{tcolorbox}

\begin{table*}[htb]
  \centering
  \caption{Comparison results between our proposed approach {\tool}  and {\tool} with different modal information}
  \begin{center}
    \begin{tabular}{c|c|ccccc}
    \toprule
    \textbf{Language} & \textbf{Approach} & \textbf{ROUGE-L (\%)}  & \textbf{METEOR (\%)}  & \textbf{BLEU-1 / 2 / 3 / 4 (\%)}
    &\textbf{CIDEr } \\   
    \midrule
    \multirow{3}{*}{Python}
    & $w/o \ desc$ & 17.73 & 12.88 & 19.08 / 12.58 / \hspace{0.3em} 9.36 / \hspace{0.1em} 7.52  & 0.60 
    \\
    & $w/o \ code$ & 27.17 & 18.64 & 29.08 / 20.46 / 15.35 / 11.99  & 1.06  
    \\
    & \textbf{SOTitle+} & \textbf{31.27} & \textbf{22.00} & \textbf{33.28 / 25.03 / 19.98 / 16.59}  & \textbf{1.43} & 
    \\
    \midrule
    \multirow{3}{*}{Java}
    & $w/o \ desc$ & 14.73 & 10.83 & 16.13 / 10.42 / \hspace{0.3em}  7.74 / \hspace{0.1em} 6.26  & 0.48  
    \\
    & $w/o \ code$ & 24.25 & 17.03 & 26.29 / 18.27 / 13.58 / 10.62  & 0.96  
    \\
    & \textbf{SOTitle+} & \textbf{27.71} & \textbf{19.46} & \textbf{30.04 / 21.99 / 17.18 / 14.01}  & \textbf{1.21} & 
    \\
    \midrule
    \multirow{3}{*}{C\#}
    & $w/o \ desc$ & 12.06 & \hspace{0.1em} 8.44 & 13.35 / \hspace{0.35em} 7.69 / \hspace{0.3em} 5.25 / \hspace{0.2em} 4.10  & 0.38  
    \\
    & $w/o \ code$ & 26.25 & 18.23 & 28.94 / 20.89  / 16.21 / 13.23  & 1.07
    \\
    & \textbf{SOTitle+} & \textbf{27.66} & \textbf{19.37} & \textbf{30.69 / 22.94 / 18.27 / 15.24}  & \textbf{1.20} 
    \\
    \midrule
    \multirow{3}{*}{JavaScript}
    & $w/o \ desc$ & 14.33 & 10.29 & 15.48 / \hspace{0.3em} 9.77 / \hspace{0.25em} 7.10 / \hspace{0.1em} 5.66  & 0.42 
    \\
    & $w/o \ code$ & 28.42 & 19.53 & 30.41 / 21.78 / 16.82  /  13.64   & 1.17
    \\
    & \textbf{SOTitle+} & \textbf{31.42} & \textbf{21.90} & \textbf{33.69 / 25.29 / 20.27 / 16.91}  & \textbf{1.41} 
    \\
    \midrule
    \multirow{3}{*}{PHP}
    & $w/o \ desc$ & 15.67 & 12.13 &  17.08  / 10.74 / \hspace{0.3em} 8.02  / \hspace{0.1em}  6.55   &  0.50 
    \\
    & $w/o \ code$ & 28.51 & 20.21  & 30.93 / 21.88  / 16.87  / 13.71  & 1.21 
    \\
    & \textbf{SOTitle+} & \textbf{31.97} & \textbf{23.20} & \textbf{34.53 / 25.80 / 20.82 / 17.59}  & \textbf{1.51}
    \\
    \midrule
    \multirow{3}{*}{HTML}
    & $w/o \ desc$ & 11.84 & \hspace{0.1em} 8.95 &  13.12  / \hspace{0.3em} 6.67  / \hspace{0.3em} 3.89  /  \hspace{0.1em} 2.59   &  0.26 
    \\
    & $w/o \ code$ & 27.64 & 20.15 & 29.66  /  19.51  /  13.69  /  10.00   &  1.02 
    \\
    & \textbf{SOTitle+} & \textbf{29.00} & \textbf{21.03} & \textbf{30.74 / 20.82 / 15.11 / 11.51}  & \textbf{1.14} 
    \\
    \bottomrule
\end{tabular}
 \end{center}
  \label{tab:RQ2}
\end{table*}%

\subsection{RQ3: Ablation Study on the Prompt-tuning Paradigm}

In this RQ, we want to investigate how much can \textbf{prompt tuning}  contribute to the performance of {\tool}. We design a control approach {\tool}$_{ft}$, which is fine-tuned on CodeT5. For {\tool}$_{ft}$, we remove all the prompts and use a special identifier ($<code>$) to distinguish problem descriptions and code snippets. The final input $X$ is defined as follows:

\begin{equation} 
\label{eq:input}
\begin{split}
   X=[LANG]\oplus X_{desc} \oplus <code> \oplus X_{code} 
\end{split}
\end{equation}
where $X_{desc}$ denotes the problem description and $x_{code}$ denotes the code snippet.

We show the comparison results between {\tool}$_{ft}$ and {\tool} in Table~\ref{tab:RQ3}. 
In this table, We find that using the prompt-tuning paradigm can outperform using the fine-tuning paradigm for all programming languages. For example, for the Python programming language, using prompt tuning can improve the performance by 14.13\%, 17.95\%, 32.00\%, and 24.47\% in terms of ROUGE-L, METEOR, BLEU-4, and CIDEr respectively. The performance improvement is particularly obvious in the low-resource scenario (i.e., programming languages with fewer question posts). For example, for the PHP programming language, prompt tuning shows the highest performance improvement, with increase rate at 13.89\%, 19.04\%, 47.94\%, and 27.97\% in terms of Rouge-L, METEOR, BLEU-4, and CIDEr, respectively. 
Finally, we also show examples on our GitHub repository\footnote{\url{https://github.com/shaoyuyoung/SOTitlePlus/blob/main/Appendix.md}} that can demonstrate the prompt-tuning contributions to bridge the gap between the fine-tuning paradigm and the prompt-tuning paradigm.

\begin{tcolorbox}[width=1.0\linewidth, title={}]
\textbf{Summary for RQ3:}
Compared with the fine-tuning paradigm, the prompt-tuning paradigm can help {\tool} achieve better performance, especially for low-resource programming languages.
\end{tcolorbox}

\begin{table*}[htb]
  \centering
  \caption{Comparison results between  {\tool} with  the prompt-tuning paradigm and {\tool} with the fine-tuning paradigm }
  \begin{center}
    \begin{tabular}{c|c|ccccc}
    \toprule
    \textbf{Language} & \textbf{Approach} & \textbf{ROUGE-L (\%)}  & \textbf{METEOR (\%)}  & \textbf{BLEU-1 / 2 / 3 / 4 (\%)}
    &\textbf{CIDEr } \\   
    \midrule
    \multirow{2}{*}{Python}
    & {\tool}$_{ft}$ & 28.53 & 20.04 & 28.88 / 21.08 / 16.52 / 13.66  & 1.24  
    \\
    & \textbf{SOTitle+} & \textbf{31.27} & \textbf{22.00} & \textbf{33.28 / 25.03 / 19.98 / 16.59}  & \textbf{1.43} & 
    \\
    \midrule
    \multirow{2}{*}{Java}
    & {\tool}$_{ft}$ & 26.36 & 18.57 & 27.52 / 19.57 / 14.86 / 11.94  & 1.09  
    \\
    & \textbf{SOTitle+} & \textbf{27.71} & \textbf{19.46} & \textbf{30.04 / 21.99 / 17.18 / 14.01}  & \textbf{1.21} & 
    \\
    \midrule
    \multirow{2}{*}{C\#}
    & {\tool}$_{ft}$ & 26.37 & 18.37 & 28.09 / 20.05 / 15.30 / 12.38  & 1.05  
    \\
    & \textbf{SOTitle+} & \textbf{27.66} & \textbf{19.37} & \textbf{30.69 / 22.94 / 18.27 / 15.24}  & \textbf{1.20} 
    \\
    \midrule
    \multirow{2}{*}{JavaScript}
    & {\tool}$_{ft}$ & 29.01 & 19.87 & 29.70 /  21.62 / 16.95 / 14.00  & 1.23 
    \\
    & \textbf{SOTitle+} & \textbf{31.42} & \textbf{21.90} & \textbf{33.69 / 25.29 / 20.27 / 16.91}  & \textbf{1.41} 
    \\
    \midrule
    \multirow{2}{*}{PHP}
    & {\tool}$_{ft}$ & 29.38 & 21.08 &  30.37  / 21.79 / 17.06  / 14.10   &  1.29 
    \\
    & \textbf{SOTitle+} & \textbf{31.97} & \textbf{23.20} & \textbf{34.53 / 25.80 / 20.82 / 17.59}  & \textbf{1.51}
    \\
    \midrule
    \multirow{2}{*}{HTML}
    & {\tool}$_{ft}$ & 27.55 & 19.68 &  29.91  /  19.67  /  13.72  /  10.16   &  1.02
    \\
    & \textbf{SOTitle+} & \textbf{29.00} & \textbf{21.03} & \textbf{30.74 / 20.82 / 15.11 / 11.51}  & \textbf{1.14} 
    \\
    \bottomrule
\end{tabular}
 \end{center}
  \label{tab:RQ3}
\end{table*}%

\subsection{RQ4: Ablation Study on Hybrid Prompts}

The prompt template design is a challenging and open problem. As \citep{liu2021gpt} reported even a single word difference in the hard prompt can make a huge performance difference. In this RQ, we want to investigate how much can our designed hybrid prompts contribute to {\tool}. 
To answer this RQ, we design two control approaches that use only hard prompts or soft prompts, respectively. We show the details of them as follows:

\begin{itemize}
    \item \textbf{Only Hard Prompts.} This control approach only uses hard prompts as prompt templates. Specifically, we use Equation (\ref{equ:hard}) as our hard prompt templates and use {$w/$ $hard$} to denote this control approach.
    
    \item \textbf{Only Soft Prompts.} This control approach only uses soft prompts as prompt templates. Specifically, we use Equation (\ref{equ:soft}) as our soft prompt templates. We use {$w/$ $soft$} to denote this control approach.
\end{itemize}

We show the performance of {\tool} with different types of prompt templates in Table~\ref{tab:RQ4}. When only using hard prompts ({$w/$ $hard$}), the performance of {\tool} has a slight performance decrease. The potential reason is that considering soft prompts in the hybrid prompt may extend the ability of prompt templates. 
It is similar to only using soft prompts ({$w/$ $soft$}) for {\tool}, which causes a slight performance decrease. The potential reason is that without original hard prompts (i.e., ``The problem description:" and ``The code snippet:"), which are the task-related important tokens, our proposed approach may lose the ability to effectively separate the problem description and the code snippet from the bi-modal information.

\begin{table*}[htb]
  \centering
  \caption{Comparison results between our proposed approach {\tool}  and the control approaches with different types of prompt templates}
  \begin{center}
    \begin{tabular}{c|c|ccccc}
    \toprule
    \textbf{Language} & \textbf{Approach} & \textbf{ROUGE-L (\%)}  & \textbf{METEOR (\%)}  & \textbf{BLEU-1 / 2 / 3 / 4 (\%)}
    &\textbf{CIDEr } \\   
    \midrule
    \multirow{3}{*}{Python}
    & $w/$ $hard$ & 30.51 & 21.19 & 32.87 / 24.23 / 19.03 / 15.99  & 1.34  
    \\
    & $w/$ $soft$ & 29.12 & 20.49 & 30.99 / 22.55 / 17.52 / 14.19  & 1.25   
    \\
    & \textbf{SOTitle+} & \textbf{31.27} & \textbf{22.00} & \textbf{33.28 / 25.03 / 19.98 / 16.59}  & \textbf{1.43} & 
    \\
    \midrule
    \multirow{3}{*}{Java}
    & $w/$ $hard$ & 27.04 & 18.89 & 29.39 / 21.19 / 16.21 / 12.99  & 1.14  
    \\
    & $w/$ $soft$ & 26.88 & 18.78 & 28.76 / 20.44 / 15.57 / 12.42  & 1.12   
    \\
    & \textbf{SOTitle+} & \textbf{27.71} & \textbf{19.46} & \textbf{30.04 / 21.99 / 17.18 / 14.01}  & \textbf{1.21} & 
    \\
    \midrule
    \multirow{3}{*}{C\#}
    & $w/$ $hard$ & 27.25 & 19.17 & 30.17 / 22.29 / 17.56 /  14.45 & 1.15 
    \\
    & $w/$ $soft$ & 26.80 & 18.76 & 29.43 / 21.50  / 16.87 / 13.87  & 1.11
    \\
    & \textbf{SOTitle+} & \textbf{27.66} & \textbf{19.37} & \textbf{30.69 / 22.94 / 18.27 / 15.24}  & \textbf{1.20} 
    \\
    \midrule
    \multirow{3}{*}{JavaScript}
    & $w/$ $hard$ & 30.76 & 21.43 & 33.18 / 24.56 / 19.42 / 16.06  & 1.34
    \\
    & $w/$ $soft$ & 29.53 & 20.53 & 31.14 / 22.72 / 17.80 /  14.68   & 1.27
    \\
    & \textbf{SOTitle+} & \textbf{31.42} & \textbf{21.90} & \textbf{33.69 / 25.29 / 20.27 / 16.91}  & \textbf{1.41} 
    \\
    \midrule
    \multirow{3}{*}{PHP}
    & $w/$ $hard$ & 31.14 & 22.76 &  33.88  / 25.02 / 19.79  / 16.39   &  1.43  
    \\
    & $w/$ $soft$ & 30.17 & 22.20  & 32.95 / 23.93  / 18.75  / 15.39  & 1.37 
    \\
    & \textbf{SOTitle+} & \textbf{31.97} & \textbf{23.20} & \textbf{34.53 / 25.80 / 20.82 / 17.59}  & \textbf{1.51}
    \\
    \midrule
    \multirow{3}{*}{HTML}
    & $w/$ $hard$ & 28.05 & 20.49 & 29.30 /  19.70  /  14.11  /  10.56    &  1.08
    \\
    & $w/$ $soft$ & 28.30 & 20.82 &  30.45  /  20.15  /  14.35  /  10.77   &  1.08 
    \\
    & \textbf{SOTitle+} & \textbf{29.00} & \textbf{21.03} & \textbf{30.74 / 20.82 / 15.11 / 11.51}  & \textbf{1.14} 
    \\
    \bottomrule
\end{tabular}
 \end{center}
  \label{tab:RQ4}
\end{table*}%

\begin{tcolorbox}[width=1.0\linewidth, title={}]
\textbf{Summary for RQ4:} 
Compared to only using hard prompts or only using soft prompts, Our results show that using hybrid prompts can achieve the best performance for {\tool}.
\end{tcolorbox}

\subsection{RQ5: Comparison with baselines via human study}
\label{sec:RQ5result}

Since CCBERT~\citep{zhang2022improving} and SOTitle~\citep{liu2022sotitle} can achieve the best performance in our considered baselines (discussed in Section~\ref{sec:resultRQ1}),
we conduct a human study to assess the quality of the question titles generated by {\tool}, CCBERT, and SOTitle respectively,  To evaluate the quality of the generated question titles, we followed a human study methodology used by the previous studies~\citep{wei2020retrieve,liu2019generating} for similar tasks.
In our human study, we consider three quality criteria (i.e., similarity, naturalness, and informativeness) and the score values range from 1 to 4, where a higher score indicates better quality of the generated question titles.
We show the meaning of these three criteria as follows:

\begin{itemize}
    \item \textbf{Similarity}. This criterion assesses the similarity between the generated question title and the ground-truth title.
    \item \textbf{Naturalness}. This criterion evaluates the grammaticality and fluency of the generated question title.
    \item \textbf{Informativeness}. This criterion quantifies the amount of content conveyed by the generated question title. Notice this criterion is irrespective of its naturalness.
\end{itemize}

We recruit six developers, including one Ph.D. student, and five Master students who are familiar with the usage of Stack Overflow and have at least three years of software development experience. They are not co-authors of this paper, which can avoid the bias of the human study results.

We randomly selected 100 posts for each programming language in the testing set, resulting in a total of 600 posts. For each post, we collected three question titles generated by {\tool}, SOTitle, and CCBERT, respectively. We split posts for six programming languages into three groups (i.e., Python and Java, C\#, and PHP, HTML, and JavaScript). Each group of posts is assigned to two different developers. Later, each developer manually evaluated 200 posts by considering the aforementioned three criteria. Notice the participants do not know which question titles are generated by {\tool}. Since the evaluation process is labor-intensive, we limit each developer to only evaluating 100 posts within a half day. Finally, the evaluation scores provided by the two developers for each post are averaged. In our human study, to improve the score quality, the recruited developers are allowed to utilize the Internet to explore unfamiliar concepts related to the posts.

\begin{figure}[htb]
	\centering
	  \includegraphics[width=0.8\textwidth]{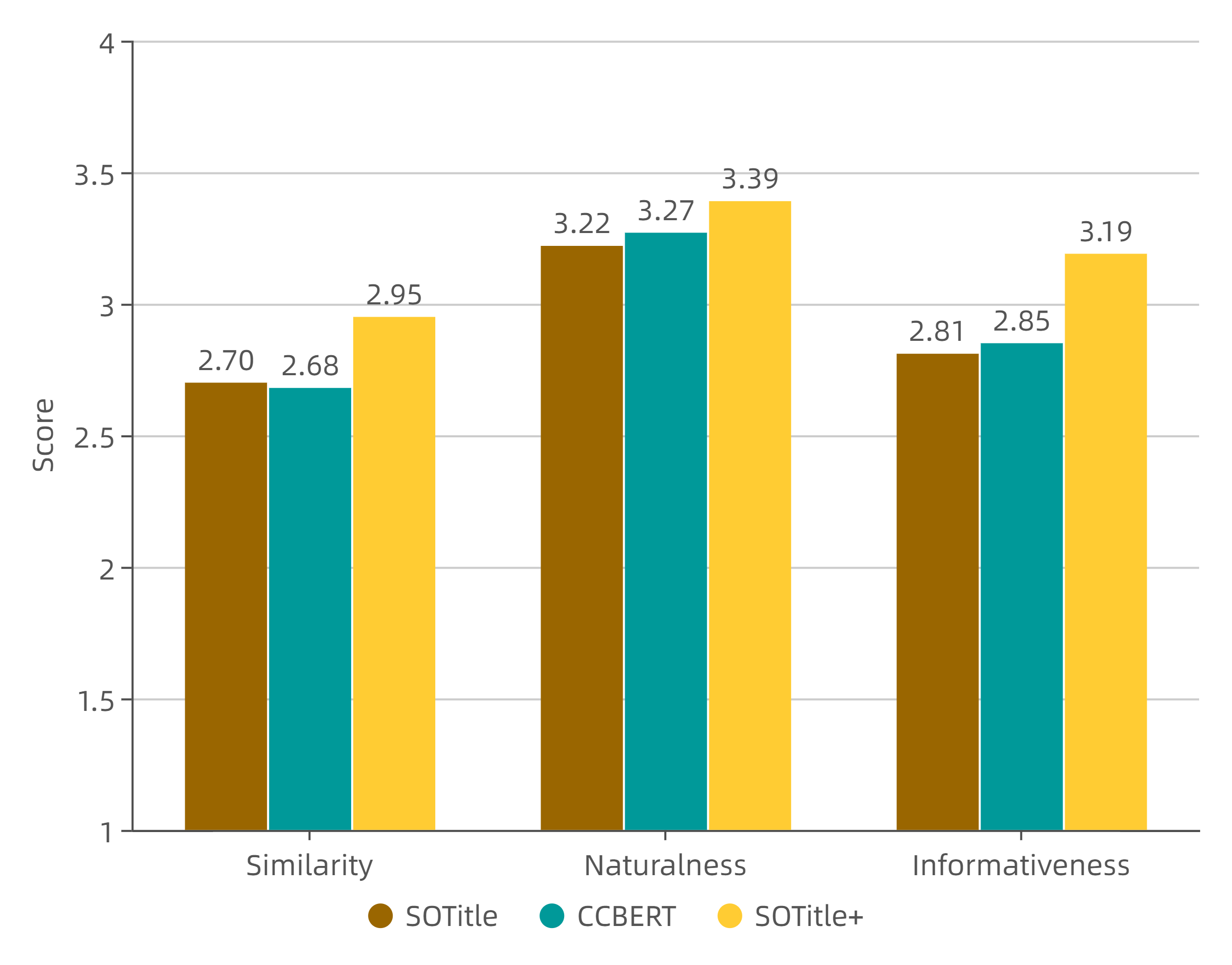}
	\caption{The average score value of our human study by evaluating similarity, naturalness, and informativeness of the generated question titles}
	\label{fig:human}
\end{figure}

The evaluation results are shown in \figurename~\ref{fig:human}. Regarding the similarity criterion, {\tool} outperforms SOTitle and CCBERT in terms of generating question titles with higher similarity scores, with an average value of 2.95. This indicates that the question titles generated by {\tool} are of good quality and can be used with minimal modifications. As for the naturalness criterion, all of these approaches achieve similar performance as expected. In terms of the informativeness criterion, {\tool} demonstrates the ability to generate more comprehensive question titles compared to SOTitle and CCBERT. To measure the consistency of the scoring results among different developers, we use Cohen’s Kappa~\citep{cohen1960coefficient}. The final Kappa value was 0.83, which indicates a substantial agreement among the developers. In addition to these scores, we also provided examples, of which {\tool} outperforms or does not outperform these two baselines, in our GitHub repository.

\begin{tcolorbox}[width=1.0\linewidth, title={}]
\textbf{Summary for RQ5:} 
Our human study shows that {\tool}
outperforms the baselines SOTitle and CCBERT by evaluating similarity,
naturalness, and informativeness of the generated question titles.
\end{tcolorbox}
\section{Discussions}
\label{sec: discuss}

\subsection{Effectiveness of Using Multi-task Learning}

In this subsection, to investigate the advantage of using multi-task learning in {\tool}, we design  $w/o \ multitask$ as the control approach. In this control approach, we train separate models for each programming language. The comparison results can be found in Table~\ref{tab:multi-task}. In this table, we find that by employing multi-task learning, {\tool}  can outperform the models trained individually for each programming language. 
For example, {\tool} achieves a performance improvement of 17.06\% at most in terms of BLEU\_4 when compared to $w/o \ multitask$ for six programming languages. Notice that using multi-task learning can achieve higher performance improvement for low-resource programming languages (such as HTML and PHP). This advantage can be attributed to the limited training data available for these low-resource languages, which can hinder the control approach $w/o \ multitask$ from effectively learning helpful knowledge. However, by utilizing multi-task learning, {\tool} can effectively acquire related knowledge from programming languages with sufficient training data.

\begin{table*}[htbp]
  \centering
  \caption{The ablation results on using multi-task learning for {\tool}}
  \begin{center}
    \begin{tabular}{c|c|ccccc}
    \toprule
    \textbf{Language} & \textbf{Approach} & \textbf{ROUGE-L (\%)}  & \textbf{METEOR (\%)}  & \textbf{BLEU-1 / 2 / 3 / 4 (\%)}
    &\textbf{CIDEr } \\   
    \midrule
    \multirow{2}{*}{Python}
    & $w/o \ multitask$ & 29.95 & 20.76 & 31.99 / 23.63 / 18.51 / 15.16  & 1.32  
    \\
    & \textbf{SOTitle+} & \textbf{31.27} & \textbf{22.00} & \textbf{33.28 / 25.03 / 19.98 / 16.59}  & \textbf{1.43} & 
    \\
    \midrule
    \multirow{2}{*}{Java}
    & $w/o \ multitask$ & 25.96 & 18.41 & 28.03 / 20.28 / 15.55 / 12.45  & 1.09  
    \\
    & \textbf{SOTitle+} & \textbf{27.71} & \textbf{19.46} & \textbf{30.04 / 21.99 / 17.18 / 14.01}  & \textbf{1.21} & 
    \\
    \midrule
    \multirow{2}{*}{C\#}
    & $w/o \ multitask$ & 25.76 & 18.13 & 28.04 / 20.48 / 16.17 / 13.41  & 1.07  
    \\
    & \textbf{SOTitle+} & \textbf{27.66} & \textbf{19.37} & \textbf{30.69 / 22.94 / 18.27 / 15.24}  & \textbf{1.20} 
    \\
    \midrule
    \multirow{2}{*}{JavaScript}
    & $w/o \ multitask$ & 29.25 & 20.73 & 31.63 / 23.19 / 18.25 / 15.00  & 1.27 
    \\
    & \textbf{SOTitle+} & \textbf{31.42} & \textbf{21.90} & \textbf{33.69 / 25.29 / 20.27 / 16.91}  & \textbf{1.41} 
    \\
    \midrule
    \multirow{2}{*}{PHP}
    & $w/o \ multitask$ & 28.98 & 21.33 &  31.95  / 23.35 / 18.35  / 15.03   &  1.30 
    \\
    & \textbf{SOTitle+} & \textbf{31.97} & \textbf{23.20} & \textbf{34.53 / 25.80 / 20.82 / 17.59}  & \textbf{1.51}
    \\
    \midrule
    \multirow{2}{*}{HTML}
    & $w/o \ multitask$ & 25.53 & 19.11 &  26.71  /  17.05  /  11.71  /  \hspace{0.1em} 8.40   &  0.90
    \\
    & \textbf{SOTitle+} & \textbf{29.00} & \textbf{21.03} & \textbf{30.74 / 20.82 / 15.11 / 11.51}  & \textbf{1.14} 
    \\
    \bottomrule
\end{tabular}
 \end{center}
  \label{tab:multi-task}
\end{table*}%

\subsection{Question Title Generation in Low-resource Scenario}
\label{sec:lowresource}

In Stack Overflow, there are some programming languages whose training data is limited and we call this scenario the low-resource scenario. 
In this subsection, we want to investigate the performance of {\tool} in this low-resource scenario. Specifically, we select other two low-resource programming languages (i.e., Ruby and Go). 
We use the same rules (introduced in Section~\ref{sec: approach}) for corresponding corpus construction and corpus split. The statistical information of these two corpora is shown in Table~\ref{tab:low-resource}.

\begin{table}[htbp]
  \centering
  \caption{Statistical information of  two corpora for low-resource programming languages (i.e., Ruby and Go)}
    \begin{tabular}{ccccc}
    \toprule
    \textbf{Language} & \textbf{Training Set} & \textbf{Validation Set} & \textbf{Testing Set} & \textbf{Total} \\
    \midrule
    Ruby  & 4,876 & 610 & 610 & 6,096\\
    Go   & 2,017  & 252 & 253 &  2,522\\
    \midrule
    \textbf{Total} & \textbf{6,893} & \textbf{862} & \textbf{863} & \textbf{8,618}\\
    \bottomrule
    \end{tabular}%
  \label{tab:low-resource}%
\end{table}%

In this experiment, we only consider two baselines (i.e., CCBERT~\citep{zhang2022improving} and SOTitle~\citep{liu2022sotitle}) due to their higher performance (discussed in Section~\ref{sec:resultRQ1}). 
For {\tool}, we directly use the model trained for answering RQ1 since we want to investigate the generalization of {\tool} on new programming languages. For the two baselines, we train SOTitle and CCBERT on corpus related to Rbuy and Go. We show the comparison results in Table~\ref{tab:few-shot}. For the Ruby programming language, {\tool} can achieve 29.98\% and 16.69\% in terms of Rouge-L and BLEU-4, respectively. Similarly, for the Go programming language, {\tool} can achieve 31.22\% and 15.02\% in terms of Rouge-L and BLEU-4, respectively. Taking the Go programming language as an example, {\tool} demonstrates the performance improvements of 13.98\%, 16.97\%, and 31.87\% in terms of the ROUGE-L, METEOR, and BLEU-4 measures, respectively when compared to SOTitle, which can achieve the best performance among these two baselines. These results verify the effectiveness of {\tool} for posts related to low-resource programming languages.

 \begin{table*}[htbp]
  \centering
  \caption{Comparison results between our proposed approach {\tool} and two baselines for two low-resource programming languages (i.e., Ruby and Go)}
  \begin{center}
  % \vspace{-2mm}
  % \setlength{\tabcolsep}{1mm}{
    \begin{tabular}{c|c|ccccc} % 添加一个"c"列
      \toprule
      \textbf{Language} & \textbf{Approach} & \textbf{ROUGE-L (\%)}  & \textbf{METEOR (\%)}  & \textbf{BLEU-1 / 2 / 3 / 4 (\%)} & \textbf{CIDEr} & \\ % 在表头添加新的列标题
      \midrule
      \multirow{3}{*}{Ruby}
      & SOTitle & 26.71 & 18.28 & 27.39 / 19.95 / 15.56 / 12.72  & 1.12 &  % 在此行末尾添加新的单元格内容
      \\
      & CCBERT & 26.85 & 14.30 & 25.79 / 17.65 / 13.14 / 10.10  & 0.93 &  % 在此行末尾添加新的单元格内容
      \\
    & \textbf{SOTitle+} & \textbf{29.98} & \textbf{21.94} & \textbf{32.04 / 24.44 / 19.83 / 16.69}  & \textbf{1.37} & 
    \\
      \midrule
      \multirow{3}{*}{Go}
      & SOTitle & 27.39 & 18.44 & 29.34 / 20.74 / 15.00 / 11.39  & 1.11 &  % 在此行末尾添加新的单元格内容
      \\
      & CCBERT & 27.47 & 15.50 & 28.50 / 19.99 / 15.18 / 11.97  & 0.99 &  % 在此行末尾添加新的单元格内容
      \\
      & \textbf{SOTitle+} & \textbf{31.22} & \textbf{21.57} & \textbf{32.80 / 23.79 / 18.45 / 15.02}  & \textbf{1.46} & % 在此行末尾添加新的单元格内容
      \\
      \bottomrule
    \end{tabular}
  \end{center}
  \label{tab:few-shot}
\end{table*}%

\subsection{Can {\tool} outperform ChatGPT?}

Recently, LLMs~\citep{brown2020language} (e.g., ChatGPT) have attracted great attention and achieved promising performance for many software engineering tasks, such as program repair~\citep{xia2023keep} and code generation~\citep{liu2023improving}.
In this subsection, we want to investigate the effectiveness of using ChatGPT in the SOQTG task and whether {\tool} can outperform ChatGPT on the posts related to Go and Ruby.
Specifically, for ChatGPT, we use OpenAI official API\footnote{\url{https://platform.openai.com/docs/api-reference}. The version of ChatGPT we use is GPT-3.5-turbo.} 
and the hard prompt introduced in Section~\ref{model-prompt-tuning} since ChatGPT does not support hybrid prompt.  
For {\tool}, we directly use the model trained for answering RQ1.  
Notice both {\tool} and ChatGPT are not trained on the corpus related to Go and Ruby. Therefore, they are compared in the zero-shot scenario. 

\begin{table*}[htbp]
  \centering
  \caption{Comparison results between our proposed approach {\tool}  and ChatGPT}
  \begin{center}
    \begin{tabular}{c|c|ccccc}
    \toprule
    \textbf{Language} & \textbf{Approach} & \textbf{ROUGE-L (\%)}  & \textbf{METEOR (\%)}  & \textbf{BLEU-1 / 2 / 3 / 4 (\%)}
    &\textbf{CIDEr } \\   
    \midrule
    \multirow{2}{*}{Ruby}
    & ChatGPT & 20.59 & 19.37 & 14.93 / \hspace{0.3em} 8.72 / \hspace{0.3em} 5.71 / \hspace{0.1em} 3.96  & 0.58
    \\
    & \textbf{SOTitle+} & \textbf{29.98} & \textbf{21.94} & \textbf{32.04 / 24.44 / 19.83 / 16.69}  & \textbf{1.37} & 
    \\
    \midrule
    \multirow{2}{*}{Go}
    & ChatGPT & 19.95 & 19.68 & 15.97 / \hspace{0.3em} 8.97 / \hspace{0.3em} 5.18 / \hspace{0.1em} 3.23  & 0.46
    \\
      & \textbf{SOTitle+} & \textbf{31.22} & \textbf{21.57} & \textbf{32.80 / 23.79 / 18.45 / 15.02}  & \textbf{1.46} & 
      \\
    \bottomrule
\end{tabular}
 \end{center}
  \label{tab:chatgpt}
\end{table*}%

\begin{figure}[htb]
	\centering
	  \includegraphics[width=0.8\textwidth]{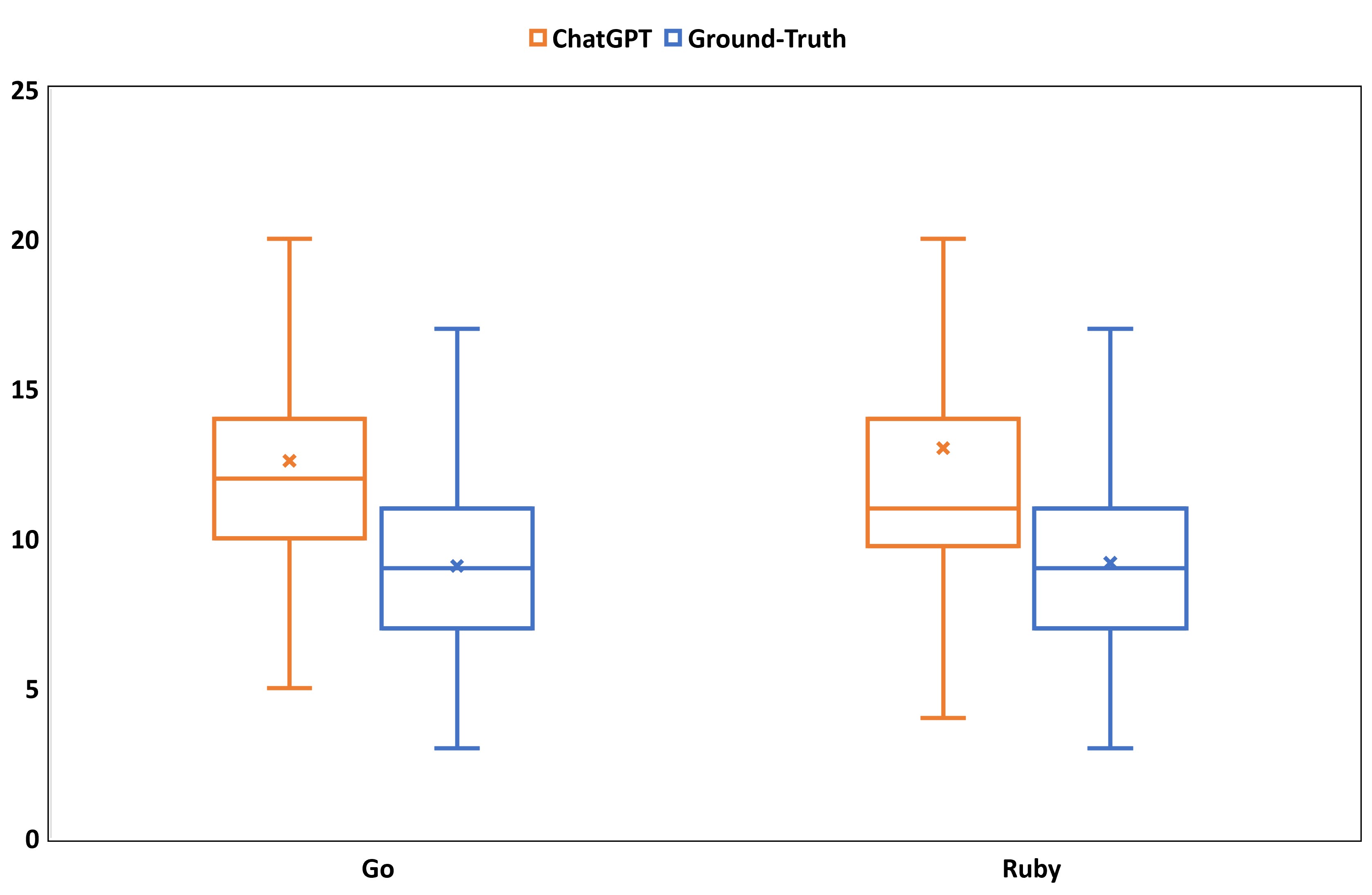}
	\caption{Length distribution of question titles generated by ChatGPT and ground-truth question titles}
	\label{fig:chatgptlength}
\end{figure}

Table~\ref{tab:chatgpt} shows the performance comparison results of ChatGPT and {\tool}. 
In this table, we can find {\tool} can achieve better performance than ChatGPT in terms of all the performance measures.
After our manual analysis, we found that ChatGPT tends to generate semantically rich but lengthy question titles.
We use Figure~\ref{fig:chatgptlength} to show the length of question titles generated by ChatGPT and ground-truth question titles. Therefore, a potential solution is to design customized prompts, which can crop the rich but lengthy generated titles, in the future.

\subsection{User Study for Improving Low-quality Question Titles}

By following the study of Code2Que~\citep{gaocode2que}, we conduct a user study to investigate whether {\tool} can generate better question titles for low-quality questions. Specifically, we sampled 20 low-quality questions for each programming language. 
Notice the votes for these questions are less than 10 and these questions are not contained in our corpus.
For each question, we gather the title written by users (i.e., ground-truth title) and the title generated by {\tool}.
Then we recruited the same six developers, who were also recruited in our human study, to determine which title was better in terms of three criteria (i.e., Clearness, Fitness, and Willingness). The detailed information of these six developers can be found in Section~\ref{sec:RQ5result}.
Specifically, Clearness evaluates the clarity of a question title. Fitness measures whether a question title is reasonable in logic with
the provided code snippet and the problem description. Willingness measures whether a user is willing to respond to this question.

We show the comparison results (i.e., Win, Loss, and Tie) for two programming languages in Table~\ref{tab:userstudy} and other programming languages in our GitHub repository. 
In this table, we find that titles generated by {\tool} can improve the title quality in terms of all three criteria in most cases.
Moreover, we find that the titles generated by {\tool} can effectively increase the willingness of users to answer questions.
Finally, we find that not all the titles of the low-quality questions can be improved.  One of the reasons is that the post itself did not provide enough information.
In summary, our preliminary user study can show the potential of {\tool} in title quality improvement.

\begin{table*}[htbp]
  \centering
  \caption{User Study ({\tool} vs. Human)}
  \begin{center}
    \begin{tabular}{c|c|ccccc}
    \toprule
    \textbf{Language} & \textbf{Evaluation Criterion} & \textbf{Win (\%)}  & \textbf{Lose (\%)}  & \textbf{Tie (\%)}\\   
    \midrule
    \multirow{3}{*}{Python}
    & Clearness & 50.83 & 30.83 & 18.33
    \\
    & Fitness & 52.50  & 25.83  & 21.67 
    \\
    & Willingness & 56.67  & 26.67  & 16.67 
    \\
    \midrule
    \multirow{3}{*}{Java}
    & Clearness & 44.17  & 39.17  & 16.67 
    \\
    & Fitness & 47.50  & 49.00  & 12.50 
    \\
    & Willingness & 39.17  & 35.83  & 25.00 
    \\
    \bottomrule
\end{tabular}
 \end{center}
  \label{tab:userstudy}
\end{table*}%

\subsection{Limitations of {\tool} by Analyzing Failed Cases}

Although {\tool} achieves promising performance via both automatic evaluation and human study, it should be noted that for some cases, {\tool} may generate question titles that show low similarity to the ground-truth question titles. To gain deeper insights into this issue, we randomly selected 100 posts and performed a manual analysis. This analysis aims to identify the specific types of posts that pose challenges for {\tool} and help to point out potential future studies.

The first limitation arises from the possibility of missing certain question words (e.g., \textit{how to, what, why}) in the generated question titles. For example, Given this question post\footnote{\url{https://stackoverflow.com/questions/51523765}}, the ground-truth title is \textit{``How to use OpenCV ConnectedComponents to get the images"} while the title generated by \tool{} is \textit{ ``OpenCV ConnectedComponents in Python"}. It appears that the model fails to capture the intention of this question post. One potential solution is to automatically identify the type of posts and provide this information for title generation.

The second challenge type is that the generated title and the ground truth convey the same meaning but are expressed using different ways. For example, given this question post\footnote{\url{https://stackoverflow.com/questions/51579215}}, the ground-truth title is \textit{``Remove seaborn lineplot legend title"}, the title generated by \tool{} is \textit{``Remove title from seaborn lineplot legend"}. The sentence-level ROUGE-L score between these two titles is 26.56\%. However, these two titles are semantically equivalent. This type of challenge can lead to lower scores in automatic evaluation metrics. One possible solution is to develop semantic-based evaluation metrics that can more accurately assess the semantic similarity between the generated title and the ground-truth title.

The third challenge type is that the problem description may contain noise and then mislead the model to generate the wrong question title. Given this question post\footnote{
\url{https://stackoverflow.com/questions/59391560}
}, the ground-truth title is \textit{``How to run UVICORN in Heroku?"}, while the title generated by {\tool} is \textit{``How to run a fastapi script in Heroku?"}. After our manual analysis, we find that the keyword \textit{``fastapi"} appears four times in the problem description and the keyword \textit{``UVICORN"} appears only one time. However, in this post, the developer wants to ask how to use a command (i.e., \textit{``UVICORN"}) but not a script (i.e., \textit{``fastapi"}). One possible solution is to design an effective approach to identify the developer's intent and use this information to guide title generation.

\subsection{Threats to validity}
\label{sec: threat}

In this section, we mainly analyze the potential threats to the validity of our empirical study.

\subsubsection{Internal Threats} 

The first threat to internal validity is related to potential faults in the implementation of our proposed approach {\tool} and baselines. To alleviate this threat, we use mature frameworks  (such as transformers, and OpenPrompt) as much as possible to implement {\tool}. We also use software testing and code inspection to guarantee the code quality of {\tool}. 
For baselines, we use their shared scripts. We replicated these scripts on the same dataset shared by these studies to confirm that they can achieve comparable results reported in the original studies. After replication, we train the models by these baselines on our gathered corpus.
The second threat to internal validity is related to prompt template construction for the SOQTG task. 
However, designing an optimal prompt template is still a challenging task due to the large-scale search space. To alleviate this threat, we considered three different types of prompt templates. Our empirical results show that using the hybrid prompt can achieve the best performance for {\tool} and outperform state-of-the-art baselines. Thus, the performance reported in our study can serve as a lower bound of {\tool}, which can be even further improved by designing more high-quality prompt templates in the future. 
The third threat to internal validity is related to the data leakage problem, which is caused by the usage of pre-trained language models. Our study used CodeT5~\citep{wang2021codet5} as the pre-trained language model. Specifically, Wang et al. employed CodeSearchNet~\citep{husain2019codesearchnet} to pre-train CodeT5, which consists of six program languages with both unimodal data and bimodal data. Moreover, they additionally considered C/CSharp from BigQuery\footnote{\url{https://console.cloud.google.com/marketplace/details/github/github-repos}}. It is not hard to find that the dataset in their pre-training task mainly comes from GitHub, while our dataset comes from Stack Overflow. Therefore, using CodeT5 as the pre-trained model can avoid the problem of dataset leakage. However, if the pre-trained model (such as PLBART~\citep{ahmad2021unified}) used by {\tool} considered Stack Overflow data, this threat exists. A practical solution is to remove the data considered in the pre-trained model from the testing set.

\subsubsection{External Threats} 

The threat to external validity concerns the quality and generalization ability of our constructed corpus. To alleviate this threat, we first consider question posts for six popular programming languages based on their popularity. Then we consider three heuristic selection rules, which aim to identify and remove low-quality posts. However, even if these rules are used, there may still exist some low-quality posts in the selected posts and more effective rules need to be designed. Finally, to alleviate the data leakage problem, we consider the temporal relationships of posts when we split the corpus, which can guarantee that the test set contains the latest posts. 
To facilitate the experiment's reproducibility, We provide our corpus process scripts on our GitHub repository.

\subsubsection{Construct Threats} 

The threat to construct validity comes from human studies and performance evaluation measures. For our human study, it may introduce evaluation bias. To alleviate this threat and ensure a fair comparison, all the developers did not know which question title is generated by {\tool} and they are not co-authors of this paper. 
Moreover, before their engagement in the study, we ensured that all the developers received a tutorial to familiarize them with the tasks and the evaluation procedures. This training aimed to minimize potential biases and inconsistencies in their assessments. For automatic evaluation, we consider four commonly used evaluation measures, which have been widely used in text-generation tasks and question title-generation studies~\citep{liu2022sotitle,hu2020deep,yang2022dualsc,el2021automatic}.
To avoid implementation errors, we use the mature framework nlg-eval to calculate these performance measure values.

\section{Related Work}
\label{sec: related}

In this section, we first summarize the related work for Stack Overflow question title generation.
Meanwhile, since we use the prompt learning paradigm in {\tool}, we also analyze the related work for using this paradigm in software engineering tasks.
Finally, we emphasize the novelty of our study.

\subsection{Stack Overflow Question Title Generation}

Ensuring the quality of question posts is essential for maintaining the usefulness of Q\&A websites (such as Stack Overflow)~\citep{yang2014asking,ponzanelli2014improving,duijn2015quality,arora2015good}. 
Recently, researchers have focused on question title generation for Stack Overflow, as high-quality titles play a significant role in attracting potential experts to provide immediate answers. \citep{gao2020generating} were the first to study this problem and proposed Code2Que, which utilizes the LSTM network with the copy and coverage mechanisms to generate question titles. 
Similar to our previous study~\citep{liu2022sotitle}, \citep{zhang2022improving} found that considering both problem description and code snippet could significantly improve the quality of generated titles. Notice this study was conducted simultaneously with our previous study.
But compared to our previous study, their study adopts a different architecture, which considers the CodeBERT model and introduces customized copy attention layers on the encoder and decoder. In addition, they simply merged the problem description and code snippet without distinguishing them with special identifiers explicitly in bi-modal construction.
Later, \citep{zhang2023diverse} further proposed the approach M3NSCT5, which utilizes the nucleus sampling and maximal marginal ranking strategy. In M3NSCT5, they fine-tuned CodeT5 and showed the promising performance of their proposed approach.
Recently, \citep{liu2023automated} further proposed an automatic question title reformulation approach QETRA by mining modification logs from Stack Overflow. Specifically,
QETRA can further polish the draft titles generated by previous question title generation approaches~\citep{gao2020generating,liu2022sotitle,zhang2022improving}.

\subsection{Prompt Learning and its Application to Software Engineering Tasks}

Prompt-tuning is a rapidly evolving technology, which directly adapts the pre-trained language model to \textit{cloze}-style prediction or sequence-to-sequence generation through some additional prompts. As a new paradigm, prompt learning has achieved promising performances on various software engineering tasks. For example, \citep{wang2022no} conducted an empirical study on three code intelligence tasks. They compared prompt-tuning with fine-tuning and reported that prompt-tuning has better performance on both classification tasks and generation tasks. \citep{zhu2023automating} proposed a context-aware prompt-tuning approach and applied this approach to two method naming tasks (i.e., method name recommendation and method name consistency checking). Recently, \citep{Li2023Vulnerability} considered various types of prompt templates and combined prompt ensembling and transfer learning to improve the performance of prediction of vulnerability characteristics.

\subsection{Novelty of Our Study}

As the example shown in \figurename~\ref{fig:motivation1}, two posts have the same code snippet, but if the problem description is different, it will result in different titles. Compared to the previous study~\citep{gao2020generating}, our study aims to generate question titles by considering both the code snippet and the problem description. Our ablation study confirms the effectiveness of considering problem description and finds the problem description contributes more to the performance of {\tool}. 
Moreover, as discussed in \figurename~\ref{fig:motivation2}, the inconsistent inputs and optimization objectives between the pre-training task and our investigated task may make fine-tuning hard to fully explore the knowledge of the pre-trained model. Therefore, {\tool} further prompt-tune CodeT5 with hybrid prompts (i.e., a mixture of hard and soft prompts) when compared to our preliminary study SOTitle~\citep{liu2022sotitle}.
The effectiveness of using prompt tuning and hybrid prompts is confirmed in our ablation studies.
Finally, to alleviate the insufficient training data issue for low-resource programming languages and fully utilize the knowledge in the related tasks, we formalize question title generation for different programming languages as separate but related tasks. Then we utilize multi-task learning to solve these tasks. The effectiveness of using multi-task learning is also confirmed in our ablation study and the promising performance is achieved when applying {\tool} to other two low-resource programming languages (i.e., Ruby and Go).
\section{Conclusion}
\label{sec: conclusion}

In this study, we propose the prompt learning-based approach {\tool} for automatic question title generation by leveraging bi-modal information (i.e., the code snippets and the problem description) in the question posts and hybrid prompts.
To verify the effectiveness of our proposed approach, we gathered 179,119 high-quality problem posts for six popular programming languages from Stack Overflow as our experimental subject.
Experimental results show the competitiveness of our proposed approach when compared with the four state-of-the-art baselines in terms of four performance measures (i.e., Rouge, METEOR, BLEU, and CIDEr).
Moreover, we also conduct a human study to verify the effectiveness of {\tool} by considering the similarity, naturalness, and informativeness of the generated question titles. 

In the future, we first want to further improve the performance of {\tool} by designing more useful prompts via prompt engineer. 
We second want to research how to fully exploit ChatGPT on the SOQTG task. The challenging task is how to crop the rich but lengthy titles that are generated by ChatGPT. Finally, We want to investigate whether vector representation of Stack Overflow posts~\citep{xu2021post2vec} can be used for {\tool} to generate question titles with higher quality.

\begin{acknowledgements}
The authors would like to thank the editors and the anonymous reviewers for their insightful comments and suggestions, which can substantially improve the quality of this work.
Shaoyu Yang and Xiang Chen have contributed equally
to this work and they are co-first authors.
Xiang Chen is the corresponding author.
This work is supported in part by 
the National Natural Science Foundation of China (Grant
no. 61202006 and 61702041) and
the Innovation Training Program
for College Students (Grant no. 2023214 and 2023356).
\end{acknowledgements}

\section*{Conflict of Interest}
The authors declare that they have no conflict of interest.

\section*{Data Availability Statements}
The datasets generated during and analyzed during the current study are available in the Github repository (\url{https://github.com/shaoyuyoung/SOTitlePlus}).


\bibliographystyle{spbasic}
% \bibliographystyle{spmpsci}      % mathematics and physical sciences
% \bibliographystyle{spphys}       % APS-like style for physics
\bibliography{bib}

\end{document}